\documentclass{kluwer} \usepackage{graphicx}
\begin{document}
\begin{article}
\begin{opening}
\title{Revisiting  the Local Scaling  Hypothesis in  Stably Stratified
Atmospheric  Boundary Layer  Turbulence: an  Integration of  Field and
Laboratory  Measurements with Large-eddy  Simulations} \author{Sukanta
\surname{Basu}\email{sukanta.basu@ttu.edu}} \author{Fernando
\surname{Port\'{e}-Agel}}   \author{Efi   \surname{Foufoula-Georgiou}}
\author{Jean-Fran\c{c}ois       \surname{Vinuesa}}      
\institute{St.   Anthony Falls  Laboratory,  University of  Minnesota,
Minneapolis, MN 55414, USA\\}

\author{Markus \surname{Pahlow}}
\institute{Lehrstuhl f\"{u}r Hydrologie, Wasserwirtschaft und 
Umwelttechnik Ruhr-Universit\"{a}t Bochum, 44780 Bochum, Germany\\}

\runningauthor{Basu  et  al.}   \runningtitle{Local  Scaling  in  the SBL}

\begin{abstract}
The  `local scaling'  hypothesis, first  introduced by  Nieuwstadt two
decades  ago, describes  the turbulence  structure of  stable boundary
layers in  a very  succinct way  and is an  integral part  of numerous
local closure-based numerical weather prediction models.  However, the
validity  of this  hypothesis under  very stable  conditions  is a
subject of on-going debate.  In  this work, we attempt to address this
controversial  issue by  performing extensive  analyses  of turbulence
data  from  several   field  campaigns,  wind-tunnel  experiments  and
large-eddy  simulations.   Wide range  of  stabilities, diverse  field
conditions and a comprehensive  set of turbulence statistics make this
study distinct.
\end{abstract}
 
\keywords{Intermittency,   Large-eddy   Simulation,   Local   Scaling,
Monin-Obukhov Similarity Theory, Stable Boundary Layer, Turbulence.}
\end{opening}

\newpage
\centerline{\bf Glossary of Symbols}
\vspace{0.5cm}
\begin{tabular}{ll}
$f_c$ &  The Coriolis  parameter\\ $g$ &  gravitational acceleration\\
$G$ &  geostrophic wind  speed\\ $H$ &  boundary layer height\\  $L$ &
Obukhov      length     ($=     -\frac{\Theta      u_*^3}{\kappa     g
(\overline{w\theta})}$)\\  $r_{mn}$ & correlation  coefficient between
$m$ and $n$\\
$u,v,w$ & velocity fluctuations (around  the average) in $x,y$ and $z$
directions\\  $U,V$   &  mean  velocity  component  in   $x$  and  $y$
directions\\      $u_*$     &      friction     velocity      (     $=
\sqrt[4]{{\overline{uw}}^2+{\overline{vw}}^2}$                      )\\
$\overline{uw},\overline{vw}$  & vertical turbulent  momentum fluxes\\
$\overline{u\theta},\overline{w\theta}$  &  longitudinal and  vertical
heat fluxes\\ $z$ & height above the surface\\ $\kappa$ & von Karman's
constant ($= 0.40$)\\
$\Lambda$ & Local Obukhov  length\\ $\sigma_m$ & standard deviation of
$m$\\  $\theta$  &  temperature  fluctuations (around  the  average)\\
$\Theta$  &  mean temperature\\  $\theta_*$  &  temperature scale  ($=
-\frac{\overline{w\theta}}{u_*}$)\\ $\zeta$  & stability parameter ($=
\frac{z}{\Lambda}$)\\
\end{tabular}

\noindent
A subscript `$_L$' on  the turbulence quantities (e.g., $u_{*L}$) will
be  used to specify  evaluation using  local turbulence  quantities --
otherwise, surface values are implied.

\section{Introduction}\label{Sec1}
In comparison  with convective and neutral  atmospheric boundary layer
(ABL)  turbulence,  stable boundary  layer  (SBL)  turbulence has  not
received much  attention despite its  scientifically intriguing nature
and practical significance (e.g., numerical weather prediction -- NWP,
and pollutant  transport).  This  might be attributed  to the  lack of
adequate   field  or  laboratory   measurements,  to   the  inevitable
difficulties in  numerical simulations  (arising from small  scales of
motion due to stratification), and to the intrinsic complexities in its
dynamics   (e.g.,  occurrences   of   intermittency,  Kelvin-Helmholtz
instability, gravity  waves, low-level jets,  meandering motions etc.)
\cite{Huntetal,Mahrt98,Derby99}.

Fortunately, the  contemporary literature is witnessing  a brisk surge
in the  SBL turbulence  research.  Field campaigns  such as  SABLES 98 
(Stable  Atmospheric Boundary-Layer Experiment in Spain 1998)
\cite{Cuxartetal}, CASES-99 (Cooperative Atmosphere-Surface Exchange
Study 1999) \cite{CASES} and high-quality wind-tunnel
experiments   \cite{Ohya97,Ohya01}    geared   towards   comprehensive
investigation of the SBL  are being carried out. In  the case of numerical
modeling,  a handful  of partially  successful  large-eddy simulations
(LESs)    were    also    attempted    during    the    last    decade
\cite{MasonDerb,Brownetal,Andren,Galmarini,
Kosovic,Saiki,Ding,Beare1}.  Very  recently, the first intercomparison
of several  LES models  for the SBL has  been conducted  as a part  of the
GABLS (Global  Energy and Water Cycle  Experiment Atmospheric Boundary
Layer  Study)  initiative   \cite{Holtslag,  Beare2}.   In the past, a  few  Direct
Numerical Simulations (DNS) of  stable shear flows were also attempted
(see \inlinecite{Barnard} and  the references therein).  However, very
low Reynolds number  ($Re \sim 10^3$) of these  simulations make their
applicability  to the ABL flows  ($Re  \sim  10^7$)  questionable.  In  a
parallel line  of research, various tools borrowed  from the dynamical
systems theory  have also been  applied to the SBL turbulence  during this
period \cite{Revelle,McNider,Basu02,vandeWiel}.

Despite  all  these  synergistic  efforts in  understanding  the  SBL,
several unresolved  (seemingly controversial) issues  still remain. It
is the purpose of this paper to address one such unresolved issue: the
validity    of    Nieuwstadt's    `local    scaling'    hypothesis
\cite{Nieuwstadt84a,Nieuwstadt84b,Nieuwstadt85,Derby90} in very stable
atmospheric boundary layers.

To achieve  this goal, we  performed extensive analyses  of turbulence
data    from   several    field   campaigns    with    diverse   field
conditions. Further  support for our  claims is provided  by analyzing
datasets from wind-tunnel experiments \cite{Ohya01} and also simulated
by  a  new  generation  LES  \cite{Porte,Porte04,Stoll,Basu}.   It  is
important  to stress  that a  combination of  statistical  analyses of
field  measurements,  laboratory data  and  numerical simulations  was
essential for  this research.  Used  in a complementary  fashion, they
increased the  reliability of  our findings by  reducing uncertainties
inherent  to  all  the   techniques.   For  instance,  in  the  stable
atmospheric  boundary layer,  presence of  mesoscale  variabilities of
unknown origin is ubiquitous.  Such mesoscale motions might complicate
the  comparisons between  observational and  theoretically anticipated
statistics.    On  the   other  hand,   information   from  controlled
wind-tunnel  experiments   are  `pristine'  in  the   sense  that  the
measurements    are   neither    subject   to    subgrid-scale   (SGS)
parameterization     errors     nor     corrupted     by     mesoscale
variabilities. However, a wind-tunnel  might never be able to simulate
the complexities  of the atmosphere  including the very  high Reynolds
number of atmospheric flows.  LES overcomes most of the aforementioned
problems but is susceptible to the SGS parameterization issues.

\section{Background}\label{Sec2}

Over land,  stable conditions are usually  characteristic of nocturnal
boundary layers  (NBLs), but  can also persist  for several  months in
polar  regions during  winter  \cite{Kosovic,Holtslag}. During  stable
stratifications,  turbulence  is  generated  by mechanical  shear  and
destroyed  by  (negative)   buoyancy  force  and  viscous  dissipation
\cite{Stull,Arya}.  This  inhibition by buoyancy force  tends to limit
the vertical extent of turbulent  mixing. It implies that the boundary
layer height ($H$)  is not an appropriate length scale  in the SBL. In his
local             scaling            hypothesis,            Nieuwstadt
\cite{Nieuwstadt84a,Nieuwstadt84b,Nieuwstadt85} conjectured that under
stable stratification  the local  Obukhov length ($\Lambda$)  based on
local  turbulent fluxes  should be  considered as  a  more fundamental
length  scale.   Then,  according  to this  hypothesis,  dimensionless
combinations of turbulent variables (gradients, fluxes, (co-)variances
etc.)  which are measured at  the same height ($z$) could be expressed
as  `universal'  functions  of   a  single  scaling  parameter  $\zeta
(=z/\Lambda)$, known as the stability parameter.  Exact forms of these
functions  could be  predicted  by dimensional  analysis  only in  the
asymptotic very stable case ($\zeta \to \infty$), as discussed below.

On clear nights with weak  winds, the land-surface becomes rather cold
due  to strong long-wave  radiative cooling and  the overlying  boundary layer
turns out  to be  very stable.  Typically,  when a surface  cools, the
heat diffusion increases and  compensates for the cooling.  But, under
very  stable  conditions,  due   to  less  efficient  vertical  mixing
associated with strong stratification, downward turbulent heat flux is
very limited --  resulting in an even colder  surface and the boundary
layer becomes  more and more  stable (a positive feedback  effect). At
some point, turbulent exchange  between the surface and the atmosphere
ceases  and the  boundary  layer becomes  decoupled  from the  surface
\cite{BelVit,Viterbo,Mahrt02}.  \inlinecite{Wyngaard}  coined the term `z-less
stratification' for  this unique decoupling phenomenon.   In this very
stable  regime, any  explicit dependence  on $z$  disappears and  as a
consequence  local  scaling   predicts  that  dimensionless  turbulent
quantities      asymptotically      approach      constant      values
\cite{Nieuwstadt84a,Nieuwstadt84b,Nieuwstadt85}.

Local  scaling  could  be  viewed  as a  generalization  of  the  well
established      Monin-Obukhov       (M-O)      similarity      theory
\cite{Monin,Sorbjan}. M-O  similarity theory is strictly  valid in the
surface  layer  (lowest 10{\%}  of  the  ABL),  whereas local  scaling
describes    the    turbulent   structure    of    the   entire    SBL
\cite{Nieuwstadt84a,Nieuwstadt84b,Nieuwstadt85}.   This means  that by
virtue of  local scaling,  field data from  the surface layer  and the
outer  layer   could  be  combined  for   statistical  analysis.   For
large-scale NWP  models with local  closure this would also  mean that
the closure scheme for the surface  layer and the outer layer could be
the same \cite{Beljaars}.

Recently, \inlinecite{Pahlow}  questioned the validity  of the concept
of  M-O similarity theory  (and thus  local scaling  hypothesis) under
very stable stratification.  Local  scaling is a powerful reductionist
approach to the SBL  \cite{Brownetal} and is an integral  part of numerous
local-closure based present-day NWP  models.  Thus, in our opinion, it
is worth to revisit and attempt to reconcile any controversy regarding
its validity.

\section{Description of Data}\label{Sec3}

We  primarily made  use  of an  extensive  atmospheric boundary  layer
turbulence dataset (comprising of fast-response sonic anemometer data)
collected by  various researchers  from the Johns  Hopkins University,
the University  of California-Davis and the University  of Iowa during
Davis   1994,  1995,  1996,   1999  and   Iowa  1998   field  studies.
Comprehensive  description of these  field experiments  (e.g., surface
cover,  fetch, instrumentation,  sampling frequency)  can be  found in
\inlinecite{Pahlow}.   We  further  augmented  this dataset  with  NBL
turbulence data from CASES-99, a cooperative  field campaign conducted near Leon, Kansas
during October  1999 \cite{CASES}. For  our analyses, data  from sonic
anemometers located at four  levels (1.5, 5, 10 and 20 m)  on the 60 m
tower  and  the adjacent  mini-tower  collected  during two  intensive
observational periods (nights of October $17^{th}$ and $19^{th}$) were
considered (the sonic anemometer at 1.5  m was moved to 0.5 m level on
October $19^{th}$).   Briefly, the collective attributes  of the field
dataset explored in this study are as follows: (i) surface cover: bare
soil, grass  and beans;  (ii) sampling frequency:  18 to 60  Hz; (iii)
sampling period: 20 to 30 minutes; (iv) sensor height ($z$): 0.5 to 20
m;  and  (v) atmospheric  stability  ($\zeta$):  $\sim0$ (neutral)  to
$\sim10$ (very stable).

The ABL field  measurements are  seldom free from  mesoscale disturbances,
wave activities, nonstationarities etc. The situation could be further
aggravated  by several kinds  of sensor  errors (e.g.,  random spikes,
amplitude  resolution error, drop  outs, discontinuities  etc.). Thus,
stringent quality control and preprocessing of field data is of utmost
importance for any rigorous  statistical analysis. Our quality control
and  preprocessing   strategies  are  qualitatively   similar  to  the
suggestions            of            \inlinecite{Vick97}           and
\inlinecite{Mahrt98b}. Specifically, we follow these steps:

(1) Visual  inspection  of individual  data  series  for detection  of
spikes,     amplitude    resolution     error,    drop     outs    and
discontinuities. Discard suspected data series from further analyses.
                                     	
(2) Adjust for changes in  wind direction by aligning sonic anemometer
data  using  60  seconds   local  averages  of  the  longitudinal  and
transverse components of velocity.

(3) Partitioning of turbulent-mesoscale  motion using discrete wavelet
transform (Symmlet-8  wavelet) with  a gap-scale \cite{Vick03}  of 100
seconds (see Figure 1  for an illustration).  Mesoscale motions (e.g.,
gravity  waves, drainage  flows)  do not  obey  similarity theory  and
should  be  removed  from  the turbulent  fluctuations  when  studying
similarity relationships \cite{Vick03}.  \inlinecite{Vick03} developed
a  Haar  wavelet  based  automated algorithm  to  detect  `co-spectral
gap-scale'  --  the  time  scale  that  separates  the  turbulent  and
mesoscale  transports.  They found  that under  near-neutral condition
the  gap-scale is  approximately  500 s  but,  sharply decreases  with
increasing stability to as low as 30 s.

Since the determination of the  gap-scales is not free from ambiguity,
in this study we decided to work  with a fixed gap-scale of 100 s. The
selection of this particular time  scale is entirely based on the past
literature usage.   Many researchers (e.g.,  Nieuwstadt 1984b, Smedman
1988, Forrer and Rotach 1997, to name a few) have long been advocating
the use  of high-pass filtering  of stably stratified  turbulence data
using a cutoff frequency of  0.01 Hz. This particular choice was based
on the evidence  of a spectral gap (minimum) at  $0.01$ Hz reported by
\inlinecite{Caughey}.  Instead  of Fourier based  high-pass filtering,
for  the  turbulent-mesoscale partitioning  we  used discrete  wavelet
transform.  The excellent localization properties of the wavelet basis
makes it a preferable candidate over the Fourier basis.

(4) Finally, to check for nonstationarities of the partitioned series,
we performed the following step:  we subdivided each series in 6 equal
intervals  and  computed the  standard  deviation  of each  sub-series
($\sigma_i,~i=1:6$).   If  $max(\sigma_i) /  min(\sigma_i)  > 2$,  the
series was discarded.

\begin{figure}
\centerline{\includegraphics[width=3.5in]{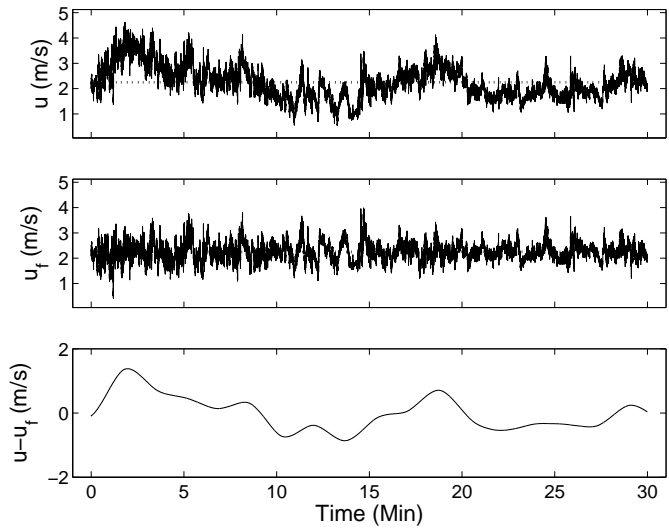}}
\caption{An  illustration  of  the  wavelet-based  turbulent-mesoscale
motion  partitioning. (Top)  longitudinal  (after alignment)  velocity
timeseries ($u$) observed during the Davis-99 field campaign; (middle)
the same velocity series ($u_f$) after wavelet filtering; and (bottom)
the mesoscale  contamination ($u - u_f$).  The  dotted line represents
the mean velocity over thirty minutes period.}
\label{Wavelet}
\end{figure}

All  the  above steps  were  performed for  all  the  3 components  of
velocity  ($u,v,w$)  and   temperature  ($\theta$),  except  that  the
nonstationarity check  (step 4) was  not performed on the  $v$ series.
This choice  was made to  ensure that we  have a sufficient  number of
runs for robust statistical analysis.  After all these quality control
and preprocessing steps  we applied, we were left  with 358 `reliable'
sets of  runs (out of  an initial total  of 633 runs) for  testing the
local scaling hypothesis.

Figure  2 portrays the  consequences of  rigorous quality  control and
preprocessing  steps on  inferences about  the validity  of  the local
scaling hypothesis.
\begin{figure}
\centerline{
\begin{tabular}{c@{\hspace{0.3pc}}c}
\includegraphics[width=2.2in]{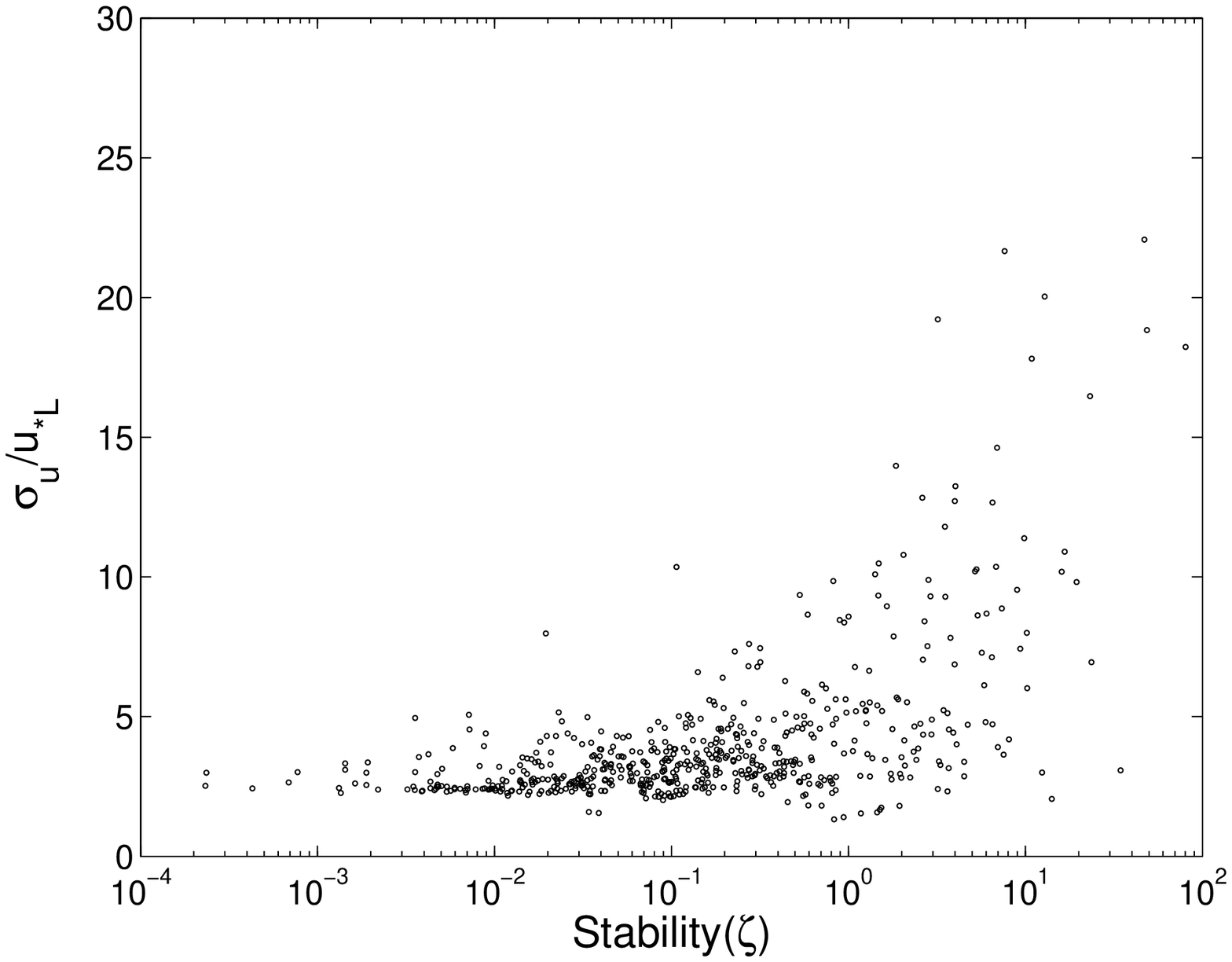}&
\includegraphics[width=2.2in]{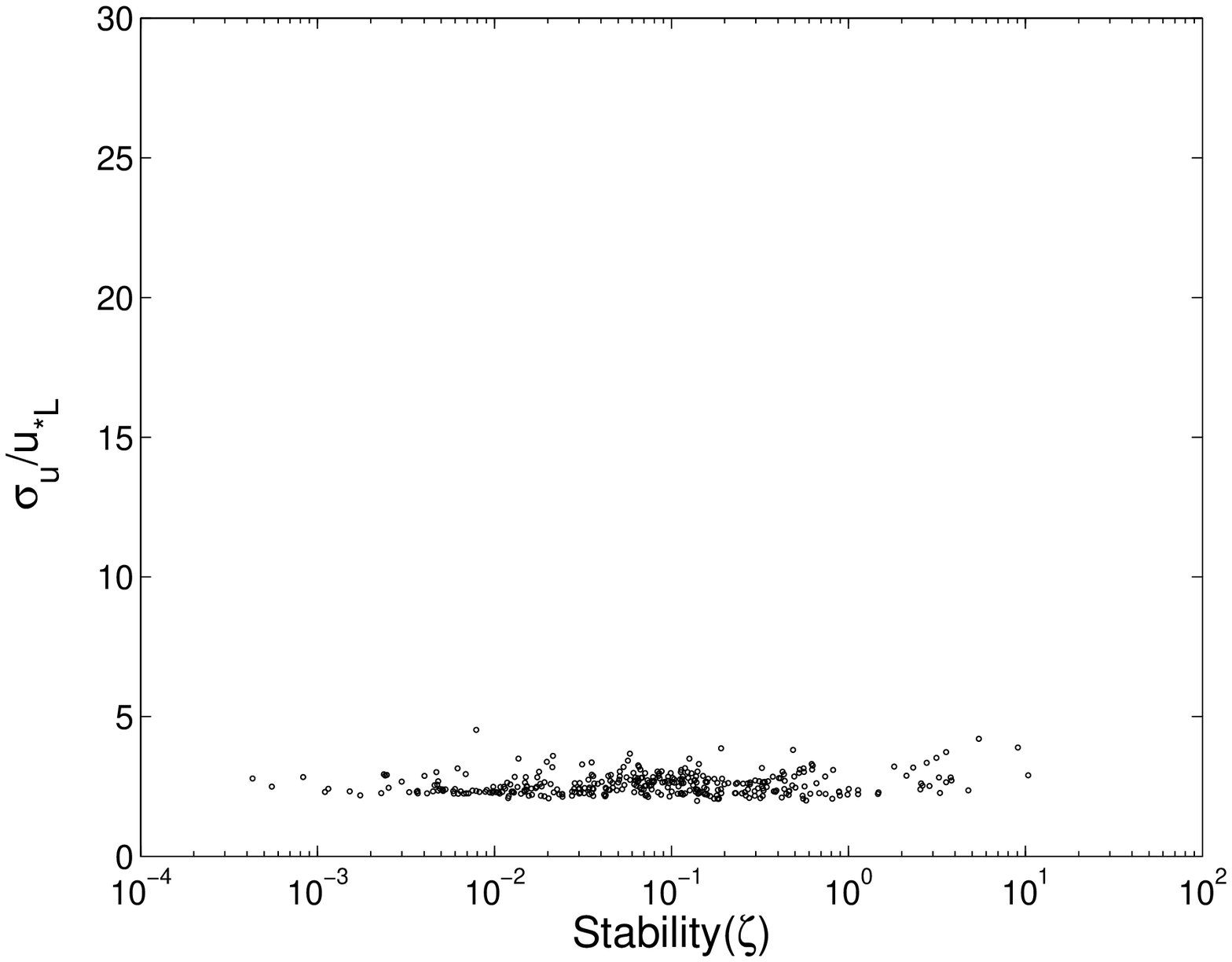}\\
\includegraphics[width=2.2in]{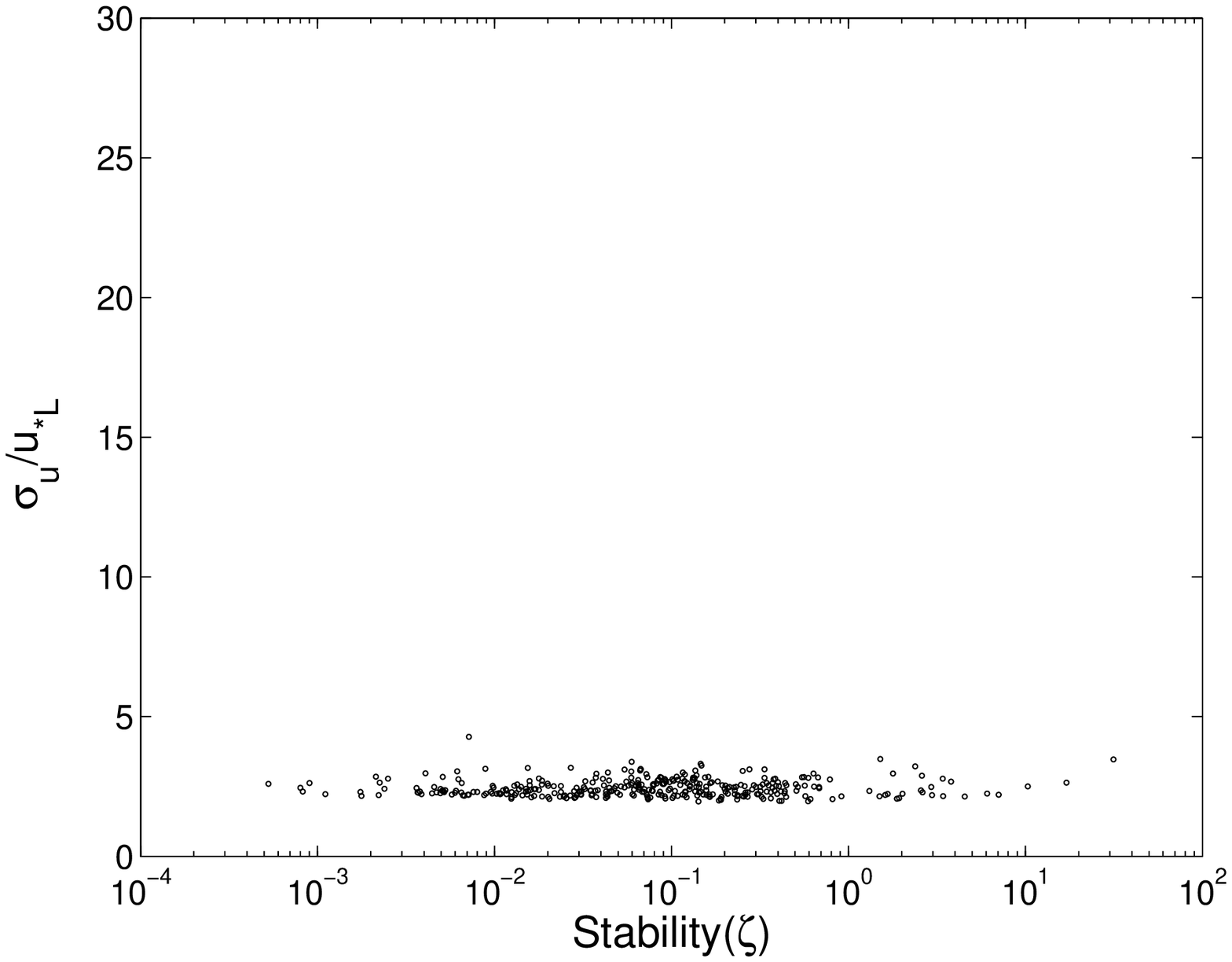}&
\includegraphics[width=2.2in]{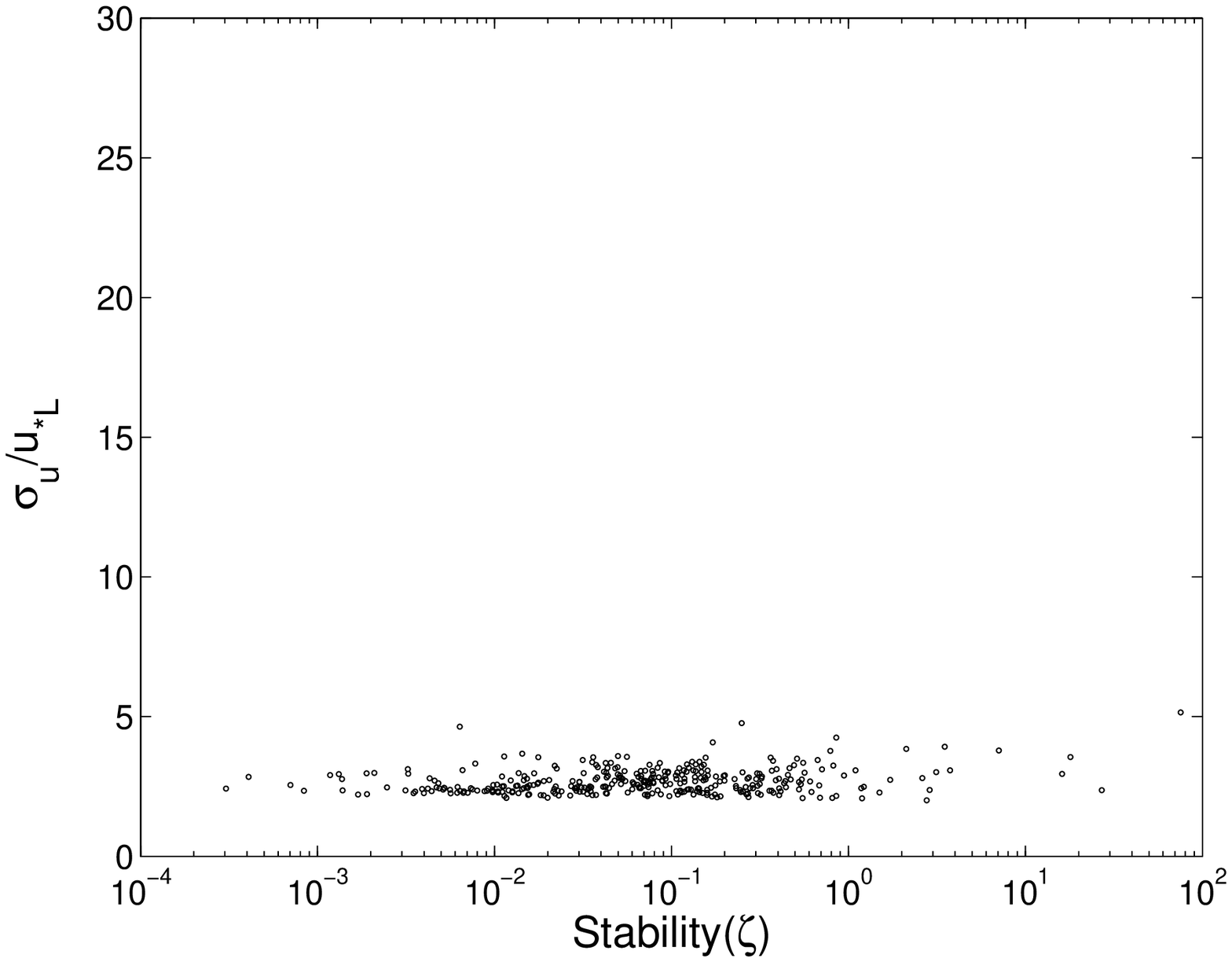}\\
\caption{$\sigma_u/u_{*L}$ versus  stability ($\zeta$) from  (top-left) field
measurements  without quality control  and preprocessing,  and (top-right)
the   same   measurements  with   appropriate   quality  control   and
preprocessing (gap-scale = 100 s). The bottom figures also correspond to the
same measurements with quality control and preprocessing but with gap-scales
of 50 s and 200 s, respectively. It is evident that the results are quite
insensitive to the range of gap-scales considered here.}
\label{Fig2}
\end{tabular}}
\end{figure}
The figure on the top-left,  representing the case without quality control
and  preprocessing (only  alignment was  done), closely  resembles the
Figure 1  of \inlinecite{Pahlow}, as  expected (since the bulk  of the
data used  in this study  were also used by  \inlinecite{Pahlow}).  On
the other hand, the figures on the top-right, bottom-left and bottom-right 
strongly supports the validity of  the local scaling  hypothesis, as  
well as  the concept  of z-less stratification.  
Later  on, in  Section \ref{Sec5} based  on extensive
analysis  of  different  sources  of  data  we  will  argue  that  the
conclusions of  \inlinecite{Pahlow} regarding the  invalidity of local
scaling and  z-less stratifications  under very stable  conditions are
biased by the inclusion of non-turbulent motions.

To substantiate this claim, we  also utilized 9 runs (corresponding to
different   levels  of   stratification)  from   the  state-of-the-art
wind-tunnel experiment by \inlinecite{Ohya01} and outputs generated by
a  new-generation LES model  \cite{Porte,Porte04,Stoll,Basu} in  conjunction with
the  field datasets.   It  is  noted that  the  field measurements  we
considered in  this study essentially represent the  surface layer; on
the other hand, the  wind-tunnel measurements and LES outputs comprise
both the  surface layer and the  outer layer.  This endows  us with an
excellent opportunity  to test the local scaling  hypothesis, since it
is supposed to  be valid for the entire  boundary layer.  However, the
influence of  boundary layer height cannot be  completely ignored near
the top  of the  boundary layer.  Note that,  the theoretical model  of Nieuwstadt
predicts singular  behavior near the boundary layer  top \cite{Nieuwstadt85}. Also
this  is the  most sensitive  location where  most of  the  LES models
considered  in the  GABLS intercomparison  differ from  each  other in
terms  of  the  blending  of  the SBL  temperature  profile  with  the
overlying inversion  \cite{Beare2}.  For these  reasons, we considered
data from the lower 75 percent  of the boundary layers (in the case of
both  wind-tunnel experiments  and LES).   Moreover, to  avoid errors
arising from flux  measurement uncertainties, wind-tunnel measurements
were further restricted such  as to satisfy the following constraints:
$u_{*L} \ge  0.  01$ m s$^{-1}$ and  $|\overline{w\theta}_L| \ge 0.001$
m K s$^{-1}$.

We  would like  to  point  out that  the  wind-tunnel measurements  of
\inlinecite{Ohya01} displayed a non-traditional upside-down character,
where, turbulence is generated in the outer boundary layer rather than
on the surface. In a recent study, \inlinecite{Mahrt02} mentioned that
even though  these boundary layers are  physically different from  the traditional
bottom-up boundary layers, the  existence of local scaling in  these boundary layers cannot be
ruled out.  Later  on in Section \ref{Sec5} we will  show that this is
indeed the case, i.e., the  local scaling and z-less features are also
found in the upside-down boundary layers.

\section{Large-Eddy Simulation of the SBL}\label{Sec4}

It  has to  be  emphasized  that the  field  observations from  stably
stratified  boundary  layers  become  increasingly uncertain  with  an
increase in stability.  This inevitable limitation highlights the need
for simulated high-resolution  spatio-temporal information about these
highly  stratified  flows to  supplement  the  observations. With  the
recent developments in  computing resources, large-eddy simulations of
turbulent flows in the ABL have  the potential to provide this kind of
information.  However, until now LES models have not been sufficiently
faithful in reproducing the characteristics of very stable atmospheric
boundary  layer \cite{Saiki,Holtslag}.   The main  weakness of  LES is
associated  with our  limited ability  to accurately  account  for the
dynamics that are not  explicitly resolved in the simulations (because
they occur at  scales smaller than the grid  size).  Under very stable
conditions -- due to  strong flow stratification -- the characteristic
size  of the  eddies  becomes increasingly  smaller  with increase  in
atmospheric stability,  which eventually imposes  an additional burden
on the  LES subgrid-scale models.   Furthermore, the recent  GABLS LES
intercomparison  study  \cite{Beare2}  highlights  that  the  LESs  of
moderately  stable  boundary layers  are  quite  sensitive  to  SGS  models  at  a
relatively fine resolution  of 6.25 m.  
At a  coarser resolution (12.5 m), occasionally, a  couple  of  traditional
SGS model-based simulations resulted in unrealistic near-linear (without
any curvature) temperature profiles. Sometimes, in these coarse-grid simulations, the 
SGS contributions to the total momentum or heat fluxes also became 
unreasonably high (much larger than fifty percent) in the interior 
of the boundary layer. These
breakdowns of traditional SGS models undoubtedly call for improved SGS
parameterizations in order  to make LES a more  reliable tool to study
very stable boundary layers.

As  a first  step  towards this  goal,  in this  study  we utilized  a
new-generation  SGS  scheme  --  the `scale-dependent  dynamic'  model
\cite{Porte,Porte04}  --  to  simulate  moderately  stable  boundary layers  at  a
relatively     coarse     resolution.      In     previous     studies
\cite{Porte,Porte04},  the  performance of  this  model in  simulating
neutral  boundary  layers  (with  passive  scalars) was  found  to  be
superior  (in  terms of  proper  near-wall  SGS dissipation  behavior,
velocity  spectra etc.)   compared to  the commonly  used  SGS models.
Technical  details  of  the  scale-dependent SGS  modeling  have  been
exhaustively described in \inlinecite{Porte} and \inlinecite{Porte04}.
To avoid repetition, we briefly  present below the basic philosophy of
this SGS modeling approach.

Eddy viscosity (eddy-diffusion) models are the most popular SGS models
in  LES of the  ABL. They  parameterize the  SGS stresses  (fluxes) as
being  proportional to the  resolved velocity  (temperature) gradients
and  involve  two  unknown  coefficients, the  so  called  Smagorinsky
coefficient  and  the  SGS   Prandtl  number.   The  values  of  these
coefficients   are  well   established   for  homogeneous,   isotropic
turbulence.   However, to  account for  shear effects  in the ABL  (due to
near-wall  effects  and   stable  stratification),  traditionally  the
eddy-viscosity   modeling  involves   appropriate   tuning  of   these
coefficients along with the use of various types of ad-hoc corrections
-- wall-damping and stability correction functions \cite{Mason}.

An alternative  approach would  be to use  the `dynamic'  SGS modeling
approach \cite{Germano,Lilly2}.  The dynamic model computes the values
of these unknown eddy-viscosity (eddy-diffusion) model coefficients at
every  time  and  locations  in  a  flow field  using  the  notion  of
scale-similarity.  Basically,  the dynamic  model avoids the  need for
{\it a-priori}  specification and consequent  tuning of any  SGS model
coefficient because it is  evaluated directly from the resolved scales
in an LES.

In  a  recent work,  by  relaxing  the  implicit assumption  of  scale
invariance  in  the   dynamic  modeling  approach,  \inlinecite{Porte}
proposed  an improved  and  more generalized  version  of the  dynamic
model:  the `scale-dependent  dynamic'  SGS model.   In  a later  work
\cite{Porte04}, the same scale-dependent dynamic procedure was applied
to estimate  the SGS scalar flux.  In  essence this procedure  not only
eliminates  the need  for any  ad-hoc assumption  about  the stability
dependence of the SGS Prandtl number but also completely decouples the
SGS  flux estimation  from  SGS stress  computation,  which is  highly
desirable.

\subsection{Description of the LES Code}
In  this  work, we  have  used  a modified  version  of  the LES  code
described    in   \inlinecite{Albertson},    \inlinecite{Porte},   and
\inlinecite{Porte04}.   The  salient  features  of this  code  are  as
follows:
\begin{itemize}[$\bullet$]
\item  It  solves  the  filtered Navier-Stokes  equations  written  in
rotational form \cite{Orszag}.
\item Derivatives in the  horizontal directions are computed using the
Fourier   Collocation   method,   while   vertical   derivatives   are
approximated with second-order central differences \cite{Canuto}.
\item Dealiasing of the nonlinear terms in Fourier space is done using
the $3/2$ rule \cite{Canuto}.
\item Explicit second-order Adams-Bashforth time advancement scheme is
used \cite{Canuto}.
\item Scale dependent dynamic SGS model with spectral cutoff filtering
is used.  The  ratio between the filter width and  grid spacing is set
to two.  The model  coefficients are obtained dynamically by averaging
locally on the horizontal plane with  a stencil of three by three grid
points  following the  approach of  \inlinecite{Zang}.  Mathematically
more  rigorous  local models  were  also  proposed  in the  literature
\cite{Piomelli,   Ghosal}. Their capabilities in the stably stratified
atmospheric boundary layer simulations have yet to be tested.

\item The scale-dependence  coefficient is determined dynamically over
horizontal      planes      following      \inlinecite{Porte}      and
\inlinecite{Porte04}.
\item Stress/flux free upper boundary condition.
\item Monin-Obukhov similarity based lower boundary condition.
\item Periodic lateral boundary condition.
\item Coriolis terms involving horizontal wind.
\item Forcing imposed by Geostrophic wind.
\item Rayleigh damping layer near the top of the domain.
\end{itemize}

\subsection{Description of Simulation}

In  this  work, we  simulated  the  GABLS  intercomparison case  study
utilizing the  scale-dependent dynamic SGS  model. This case  study is
described in  detail in Beare  et al.  (2005).  Briefly,  the boundary
layer  is driven  by an  imposed, uniform  geostrophic wind  ($G  = 8$
m s$^{-1}$),  with a  surface cooling  rate of  $0.25$ K  per  hour and
attains a quasi-steady state in $\sim$ 8-9 hours with a boundary layer
depth of  $\sim 200$  m.  The initial  mean potential  temperature was
$265$ K  up to $100$ m with  an overlying inversion  of strength $0.01$
K m$^{-1}$.  The  Coriolis parameter  was set to  $f_c = 1.   39 \times
10^{-4}$  s$^{-1}$, corresponding  to latitude  $73^o$ N.   Our domain
size was: ($L_x = L_y = L_z  = 400$ m).  This domain was divided into:
(1) $N_x \times N_y \times N_z =  32 \times 32 \times 32$ nodes (i.e.,
$\Delta_x = \Delta_y = \Delta_z = 12.5$ m); (2) $N_x \times N_y \times
N_z =  64 \times  64 \times  64$ nodes (i.e.,  $\Delta_x =  \Delta_y =
\Delta_z = 6.25$ m); and (3) $N_x \times N_y \times N_z = 80
\times 80 \times 80$ nodes (i.e.,  $\Delta_x = \Delta_y = \Delta_z = 5$ m). 
One of the objectives  behind these simulations was to investigate
the sensitivity of our results on grid-resolution.

The lower boundary condition  is based on the Monin-Obukhov similarity
theory.   The   instantaneous  wall  shear   stress  $\tau_{i3,w}$  is
represented as  a function of  the resolved velocity  $\tilde{u}_i$ at
the grid point immediately above the  surface (i.e., at a height of $z
= \Delta_z/2$ in our case):
\begin{equation}
\tau_{i3,w} = -u_*^2\left[\frac{\tilde{u}_i(z)}{U(z)}\right]
\hspace{0.25in} (i=1,2)
\end{equation}
where $u_*$ is the friction  velocity, which is computed from the mean
horizontal   mean   velocity   $U(z)   =  \langle   (\tilde{u}_1^2   +
\tilde{u}_2^2)^{1/2}  \rangle$   at  the  first  model   level  ($z  =
\Delta_z/2$) as follows:
\begin{equation}
u_* = \frac{U(z)\kappa}{\log(\frac{z}{z_o})+\beta_m \frac{z}{L}}
\end{equation}
In a similar manner, the heat flux is computed as:
\begin{equation}
\overline{w\theta}    =     \frac{u_*    \kappa    \left[\theta_s    -
\Theta(z)\right]}{\log(\frac{z}{z_o})+\beta_h \frac{z}{L}}
\end{equation}
where $\theta_s$  and $\Theta(z)$  denote the surface  temperature and
the  mean resolved  potential temperature  at the  first  model level,
respectively.    Following   the    recommendations   of   the   GABLS
intercomparison study, the constants  $\beta_m$ and $\beta_h$ were set
to 4.8 and 7.8, respectively.

\section{Results}\label{Sec5}

The mean profiles of wind speed, potential temperature, momentum flux,
and heat flux averaged over  the final hour (8-9 hours) of simulation,
are  shown in Figure  3.  The  shapes and  features of  these profiles
(e.g., super-geostrophic  nocturnal jet near  the top of  the boundary
layer, linear  heat flux profile) are in  accordance with Nieuwstadt's
theoretical   model   for    `stationary'   stable   boundary   layers
\cite{Nieuwstadt85}  and  also  very  similar to  the  fine-resolution
simulations   described  in  the   GABLS  LES   intercomparison  study
\cite{Beare2}.

\begin{figure}
\centerline{
\begin{tabular}{c@{\hspace{0.3pc}}c}
\includegraphics[width=2.2in]{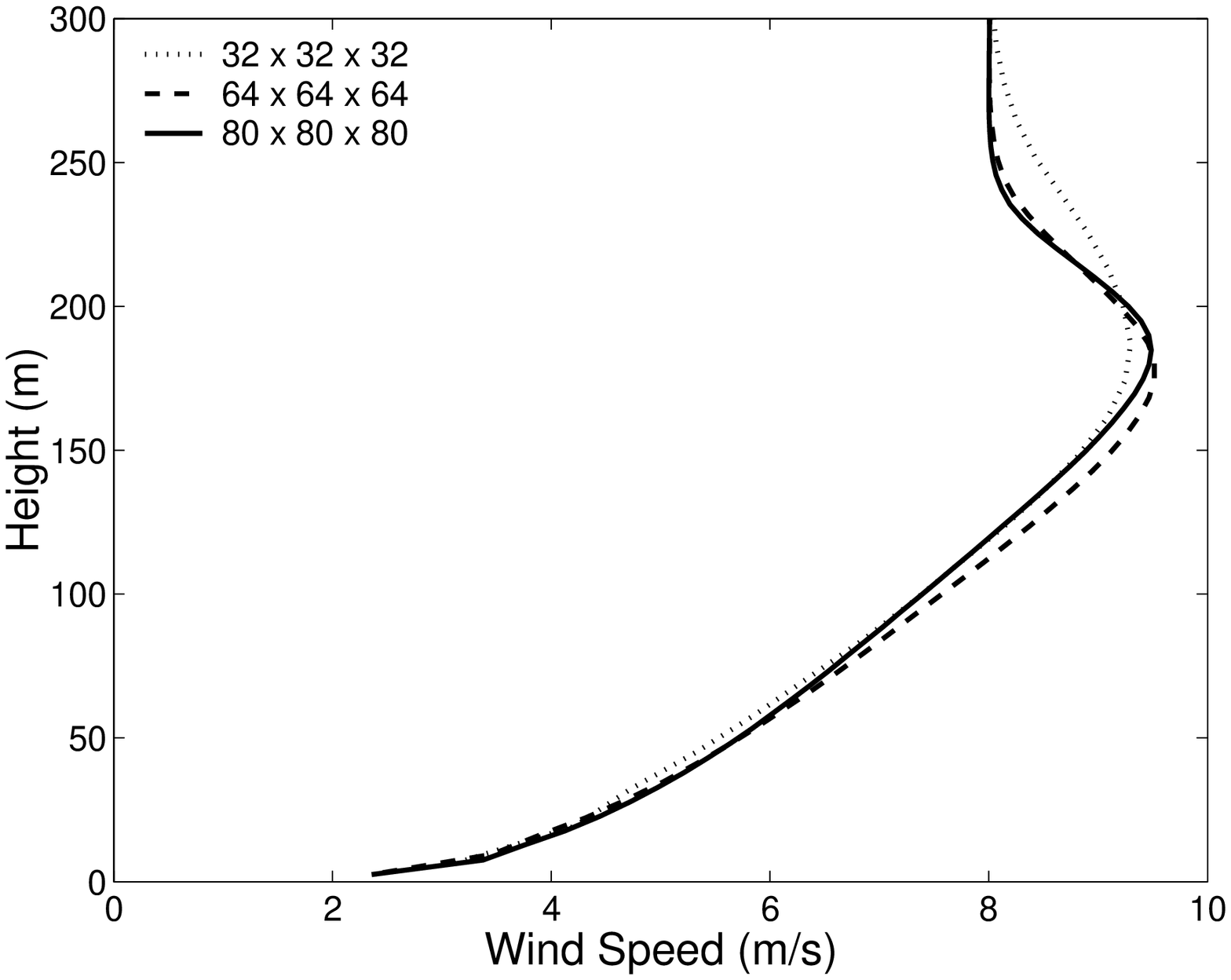}&
\includegraphics[width=2.2in]{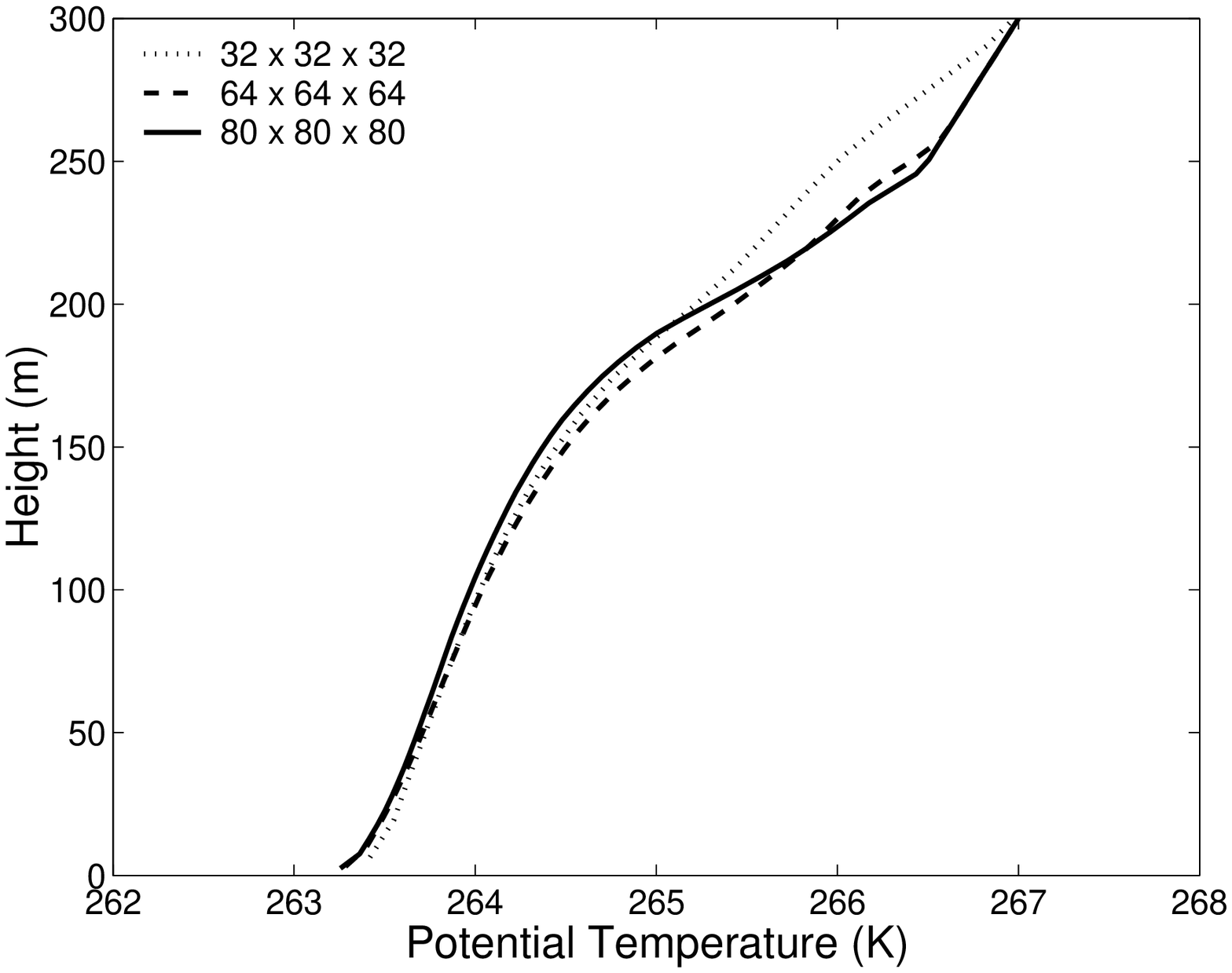}\\
\includegraphics[width=2.2in]{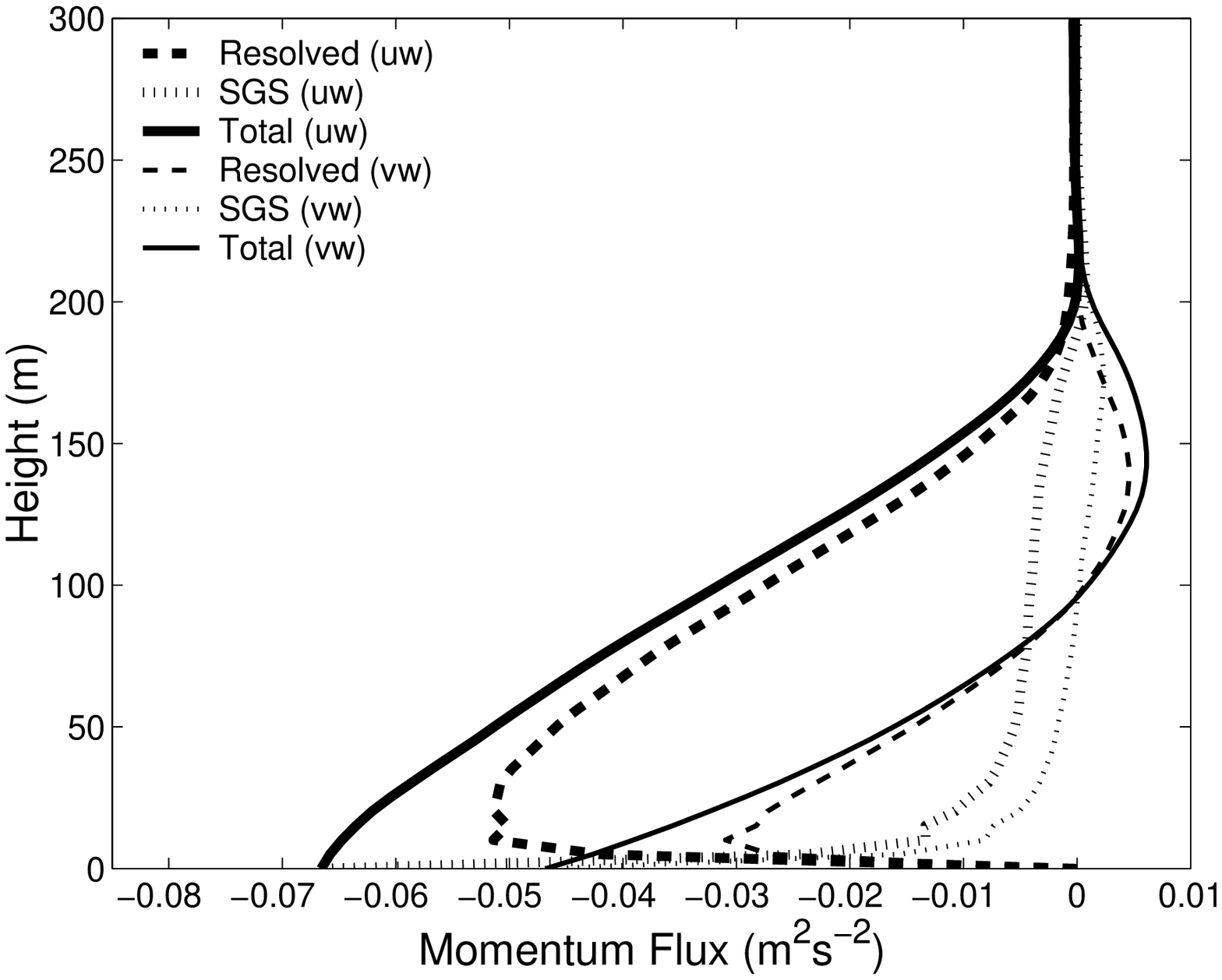}&
\includegraphics[width=2.2in]{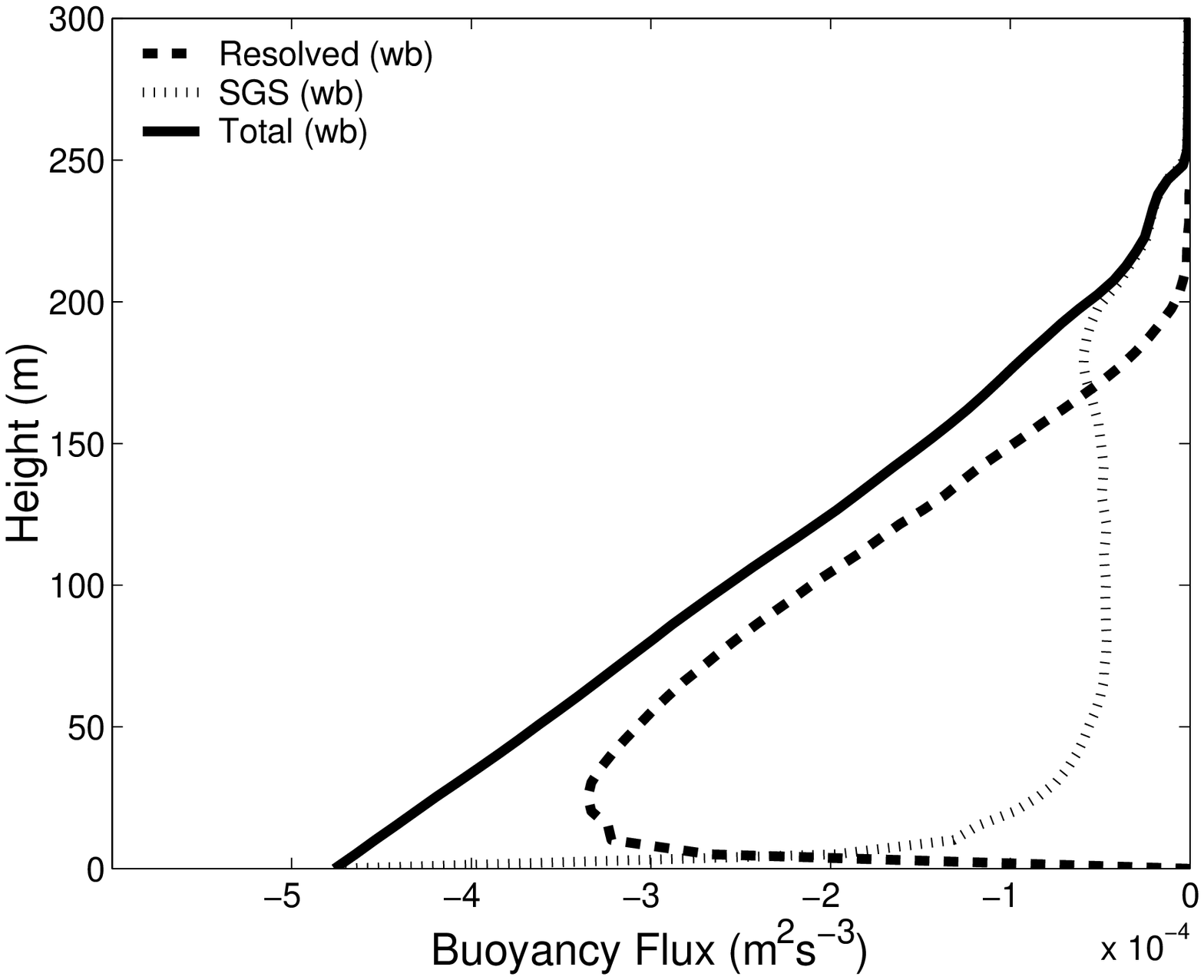}\\
\caption{Mean wind speed (top-left) and potential temperature profiles
(top right)  for four different  resolutions.  (Bottom-left) momentum
flux  and (bottom-right)  heat  flux profiles  correspond  to the  $80
\times 80 \times 80$ simulation.  These profiles are averaged over the
last one hour of simulation.}
\label{GABLS}
\end{tabular}}
\end{figure}

The  boundary  layer height\footnote{Following  \cite{Kosovic,Beare2},
the boundary  layer height is  defined as $(1/0.95)$ times  the height
where  the mean  local stress  falls to  five percent  of  its surface
value.} ($H$),  Obukhov length ($L$) and other  characteristics of the
simulated SBLs (averaged over the  final hour of simulation) are given
in Table  1. From this table  and also from Figure  \ref{GABLS}, it is
apparent that the simulated (bulk) boundary-layer parameters are quite
insensitive to  the grid-resolution.  In  LES this behavior  is always
desirable and its existence is usually attributed to the strength of a
SGS model. Whether or  not the simulated turbulence statistics support
the  local scaling  hypothesis  will be  discussed  shortly.  The  LES
statistics are computed from the  last one hour of the simulation. All
the  LES  statistics are  computed  in  the  original model  frame  of
reference.   Small   corrections  due  to  wind   rotation  have  been
neglected.  In the scale-dependent dynamic modeling approach, one does
not solve additional prognostic equations for the SGS turbulence 
kinetic energy (TKE) and the SGS scalar variance. Thus, in order to estimate the  SGS contributions 
to the total standard deviations, we followed the  approach  of   Mason
\cite{Mason89,MasonDerb}. 

\begin{table}
\caption{Basic characteristics  of the simulated SBLs  during the last
hour of simulation}\label{T1}
\begin{tabular}{ccccc} \hline
Grid Points &  $h$ (m) & $L$ (m) & $u_*$  (m s$^{-1}$) & $\theta_*$ (K)
\\ \hline  
$32\times32\times32$ & 205 & 113  & 0.283 & 0.047  \\ 
$64\times64\times64$ & 185 & 114  & 0.276 & 0.045  \\
$80\times80\times80$ & 192 & 122  & 0.285 & 0.045  \\
\hline
\end{tabular} 
\end{table}

For ease in representation, we categorize our entire database based on
local stabilities  ($z/\Lambda$) (see  Table \ref{T2}).  The  class S1
represents near  neutral stability; while  S5 corresponds to  the very
stable regime. We would like to point out that most of the very stable 
samples in the large-eddy simulations come from the interior of the 
boundary layer, rather than the surface layer. This is quite advantageous 
since the influences of the SGS terms significantly diminish away from 
the surface layer.     

\begin{table}
\caption{Number of samples in each stability class}\label{T2}
\begin{tabular}{ccccc} \hline
Class & Stability & Field &  Wind Tunnel & Large-Eddy \\ & ($\zeta$) &
Observations &  Measurements & Simulations\\  \hline S1 &  0.00-0.10 &
200 & 15 & 3 \\ S2 & 0.10-0.25 &  70 & 11 & 6 \\ S3 & 0.25-0.50 & 41 &
24 & 8 \\ S4 & 0.50-1.00 & 23 & 20 & 11 \\ S5 &$>$~1.00 & 24 & 7 & 33
\\
\hline
\end{tabular} 
\end{table}

\begin{figure}
\centerline{\includegraphics[width=2.5in]{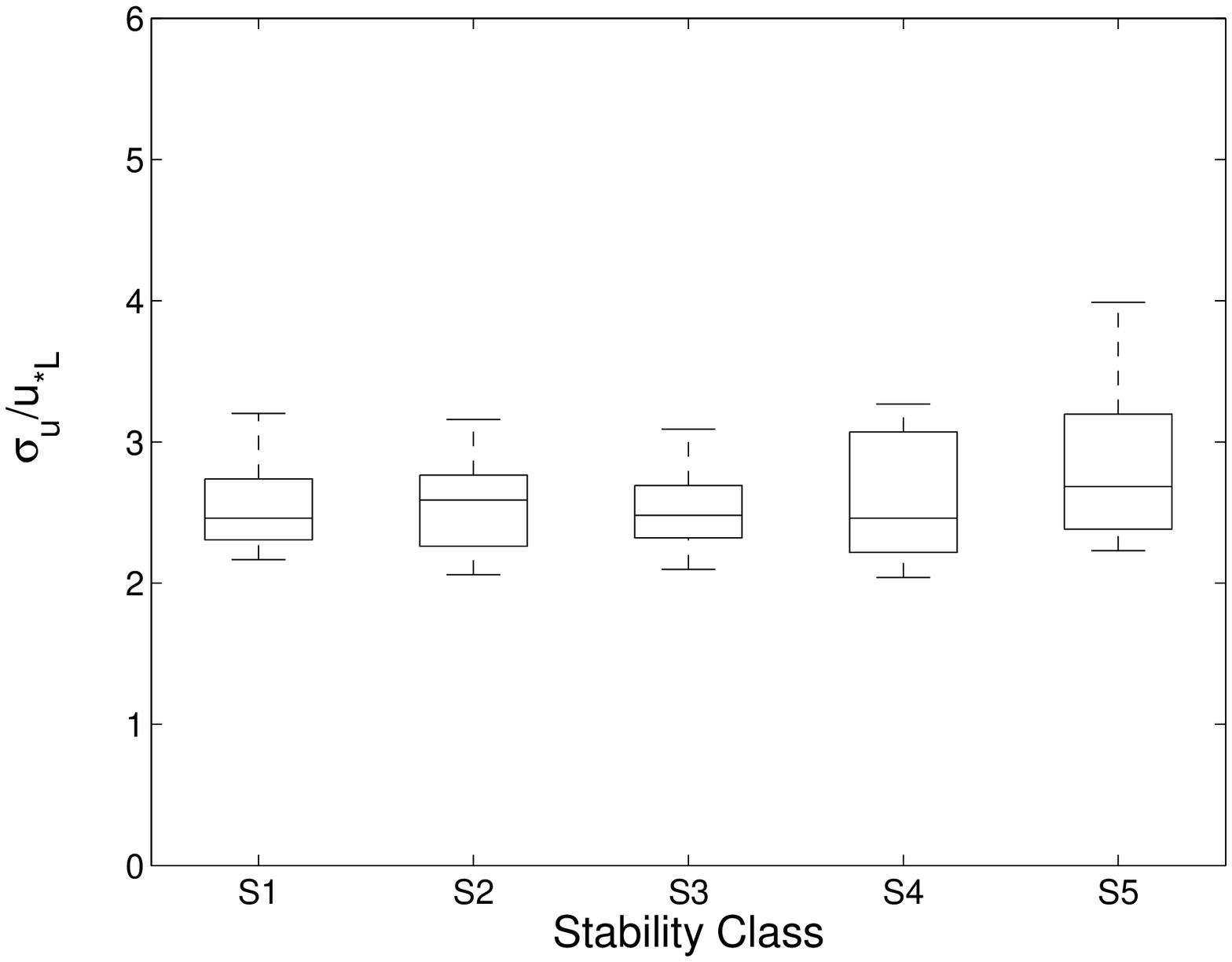}}
\centerline{\includegraphics[width=2.5in]{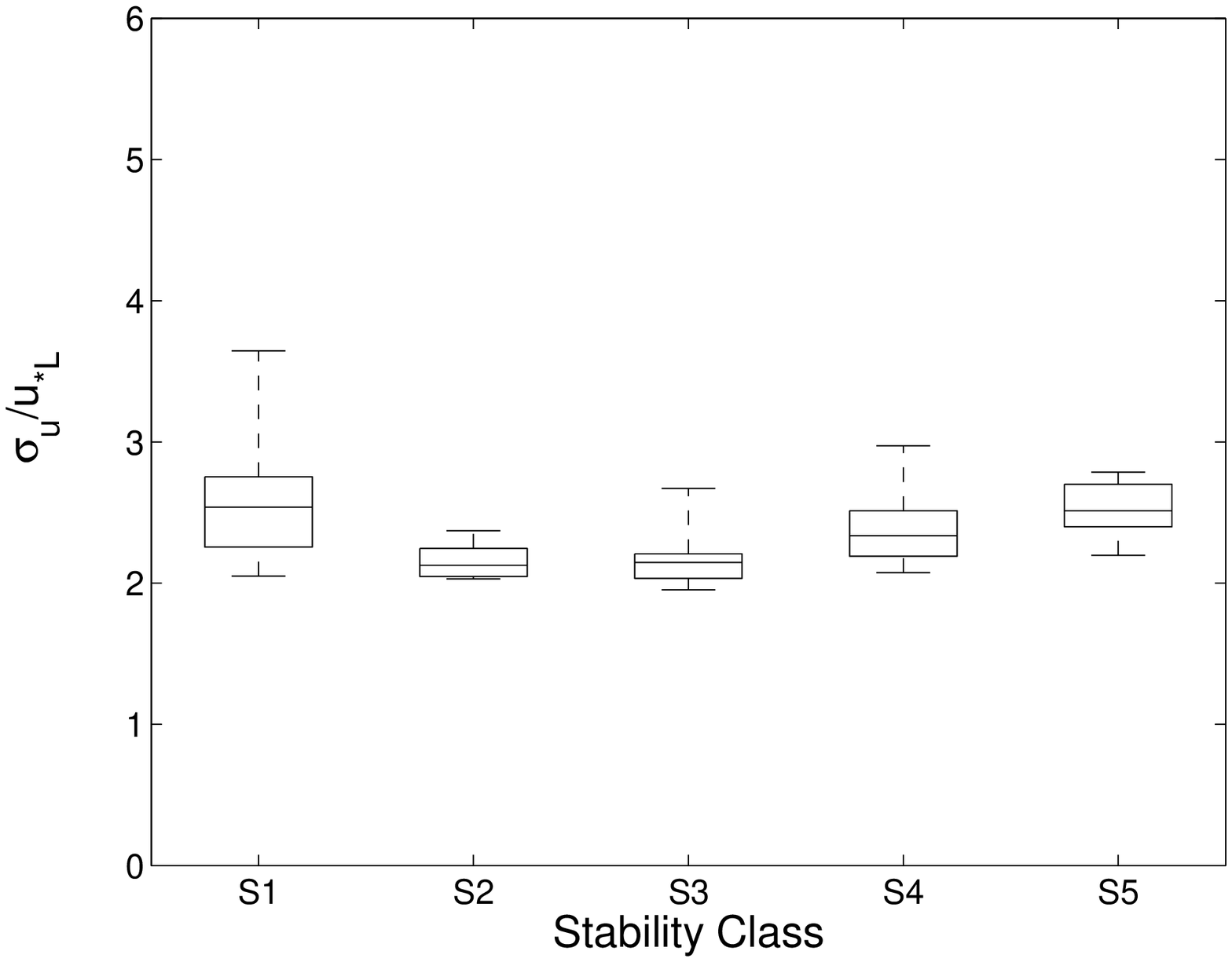}}
\centerline{\includegraphics[width=2.5in]{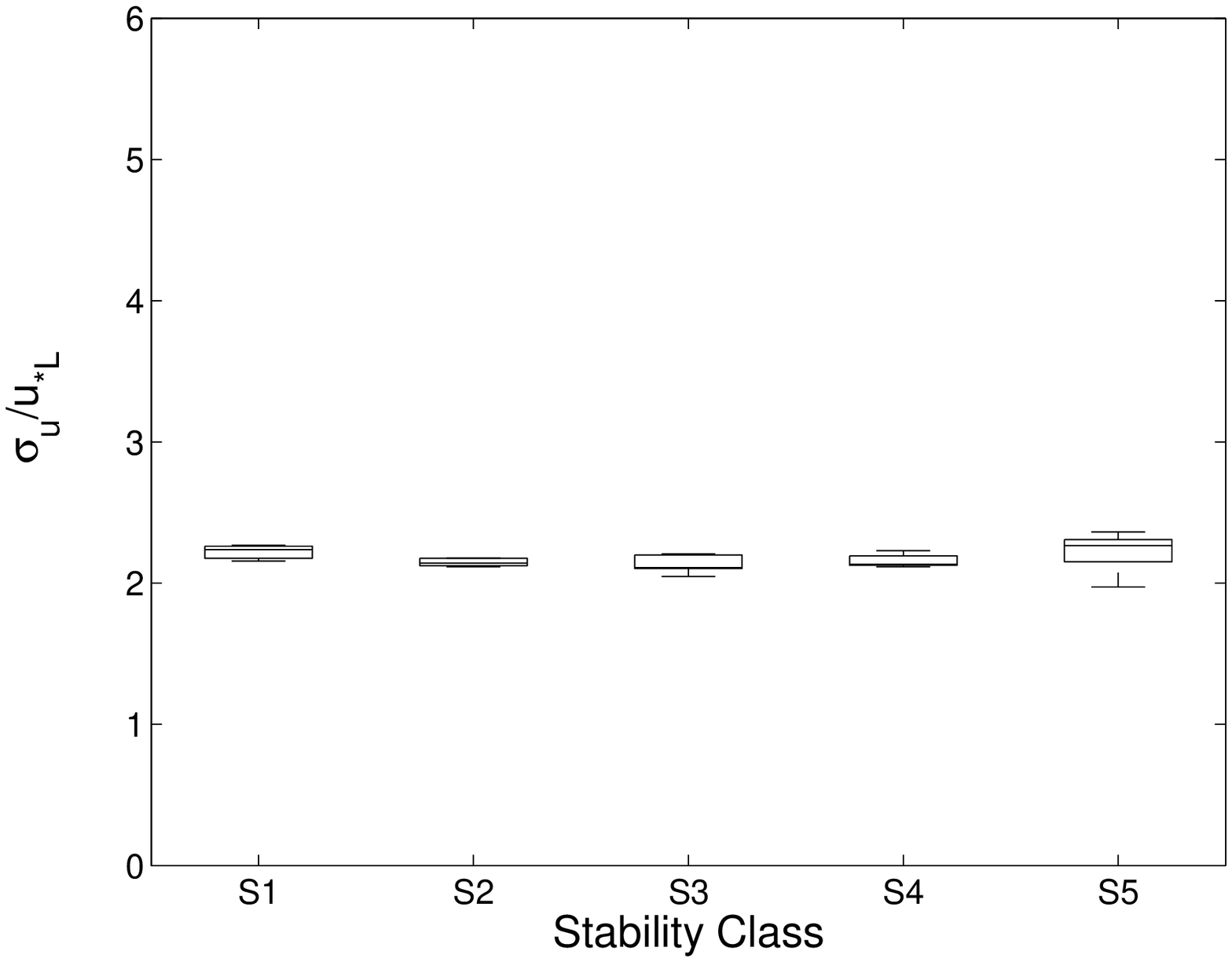}}
\caption{$\sigma_u/u_{*L}$   from  (top)  field   measurements,  (middle)
wind-tunnel measurements, and (bottom) large-eddy simulations.}
\label{SigU}
\end{figure}

\begin{figure}
\centerline{\includegraphics[width=2.5in]{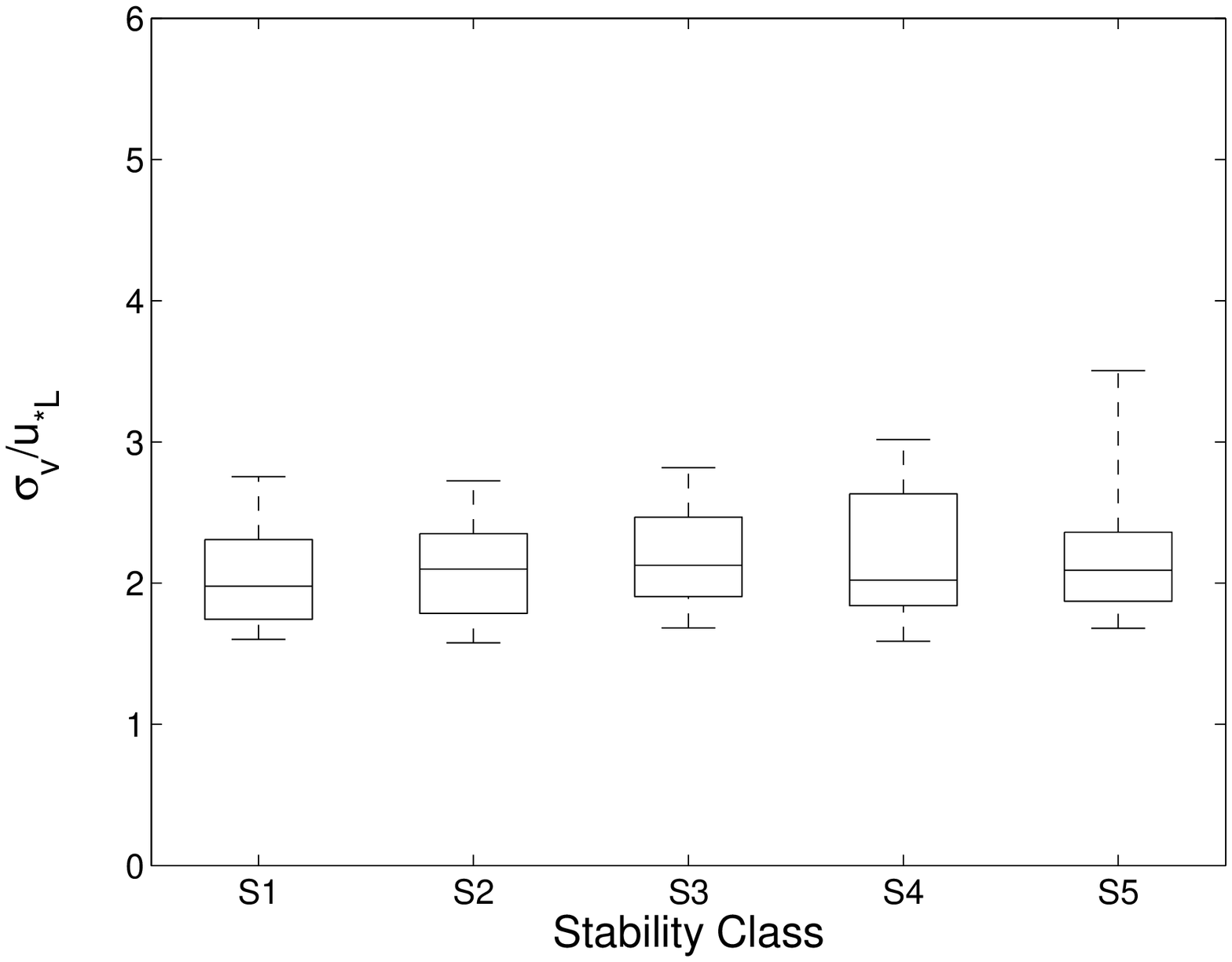}}
\centerline{\includegraphics[width=2.5in]{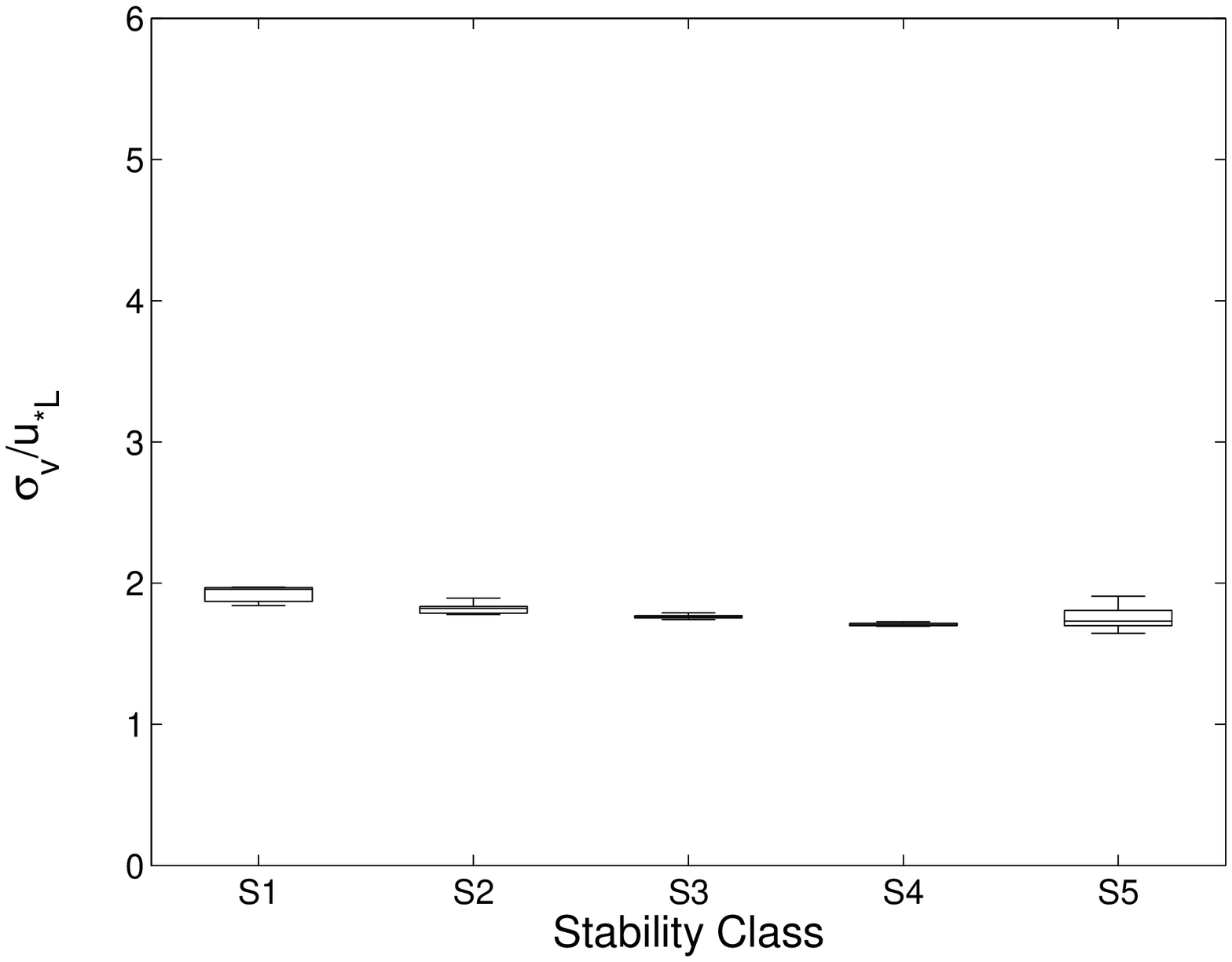}}
\caption{$\sigma_v/u_{*L}$  from (top)  field measurements,  and (bottom)
large-eddy simulations.}
\label{SigV}
\end{figure}

\begin{figure}
\centerline{\includegraphics[width=2.5in]{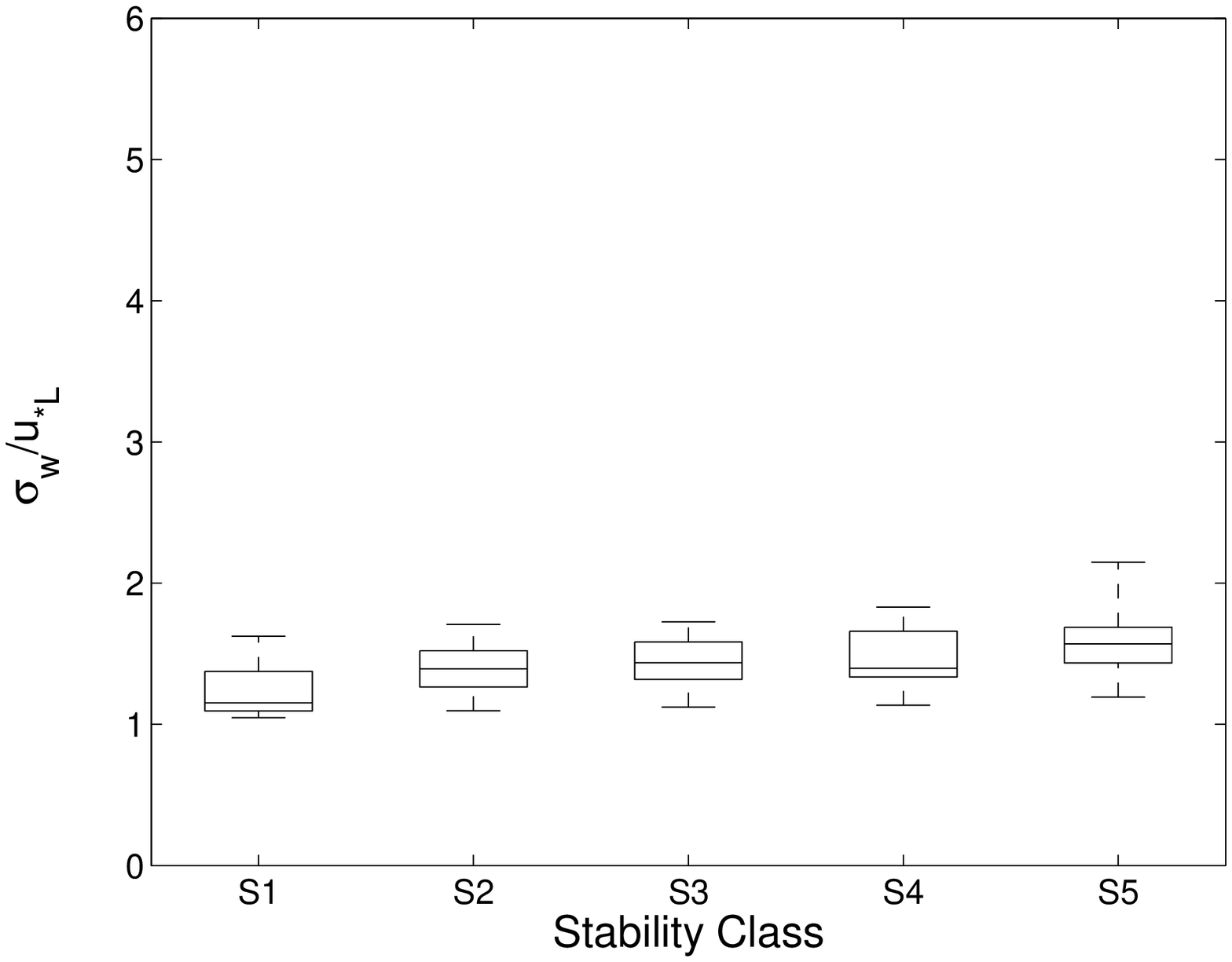}}
\centerline{\includegraphics[width=2.5in]{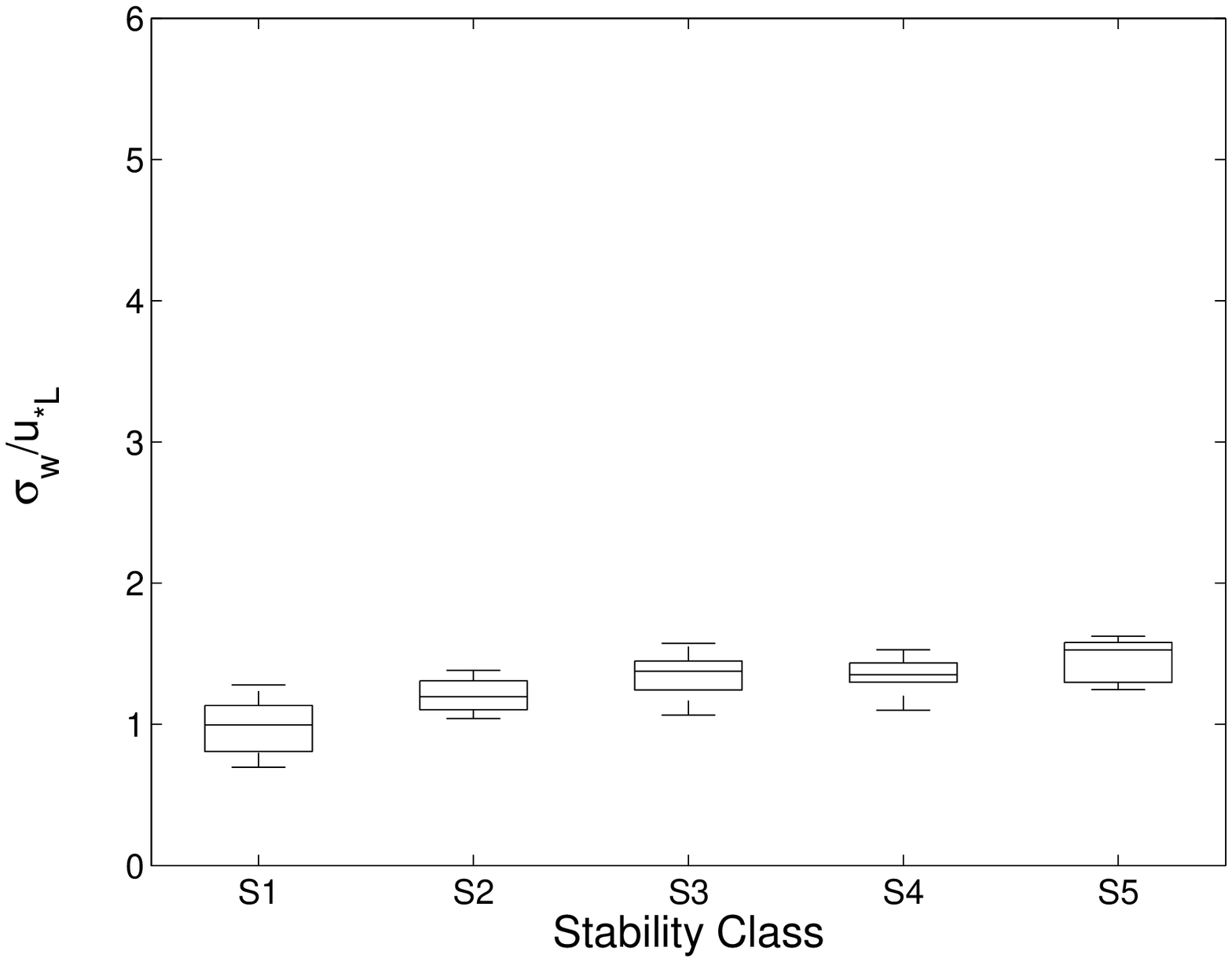}}
\centerline{\includegraphics[width=2.5in]{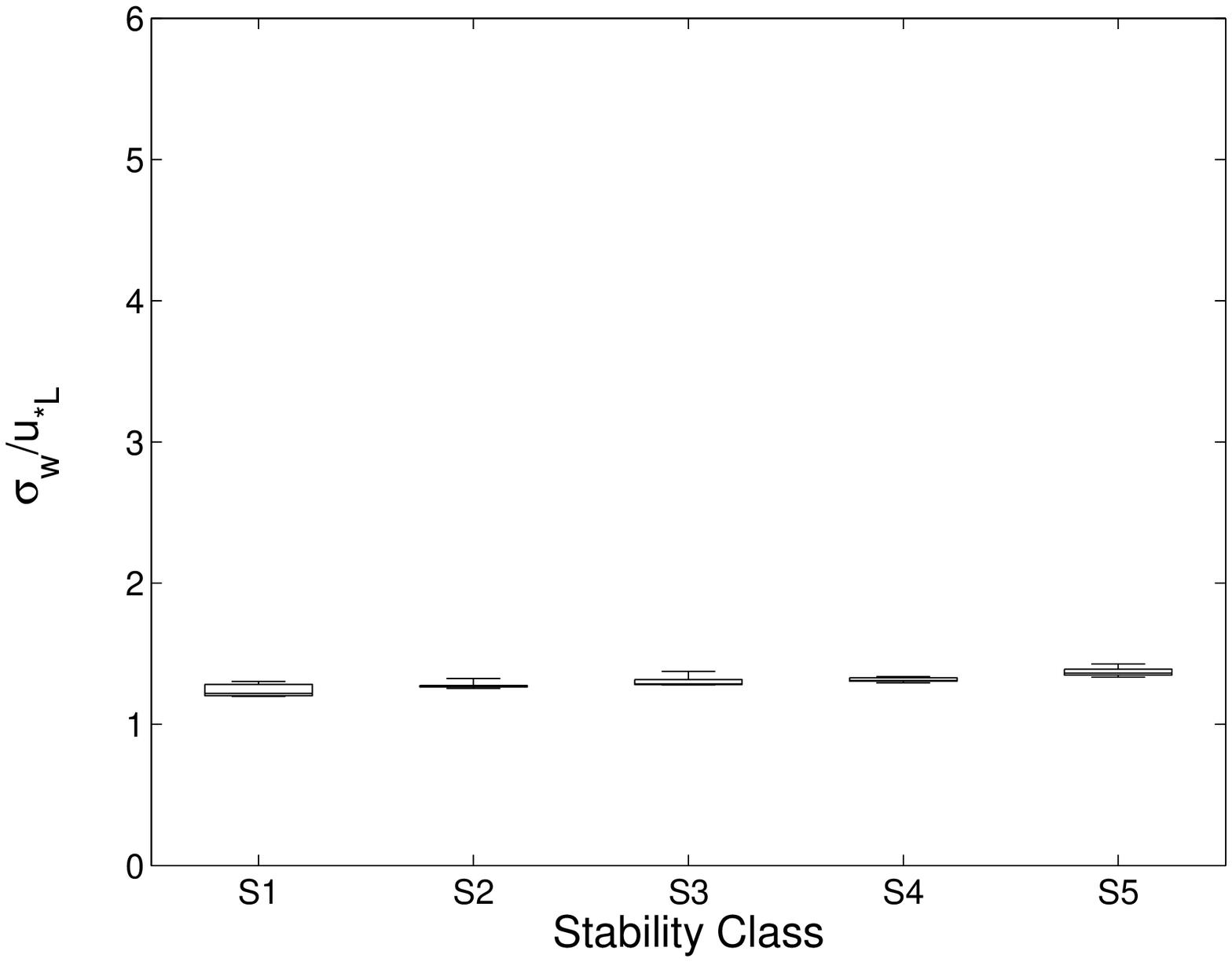}}
\caption{$\sigma_w/u_{*L}$   from  (top)  field   measurements,  (middle)
wind-tunnel measurements, and (bottom) large-eddy simulations.}
\label{SigW}
\end{figure}

\begin{figure}
\centerline{\includegraphics[width=2.5in]{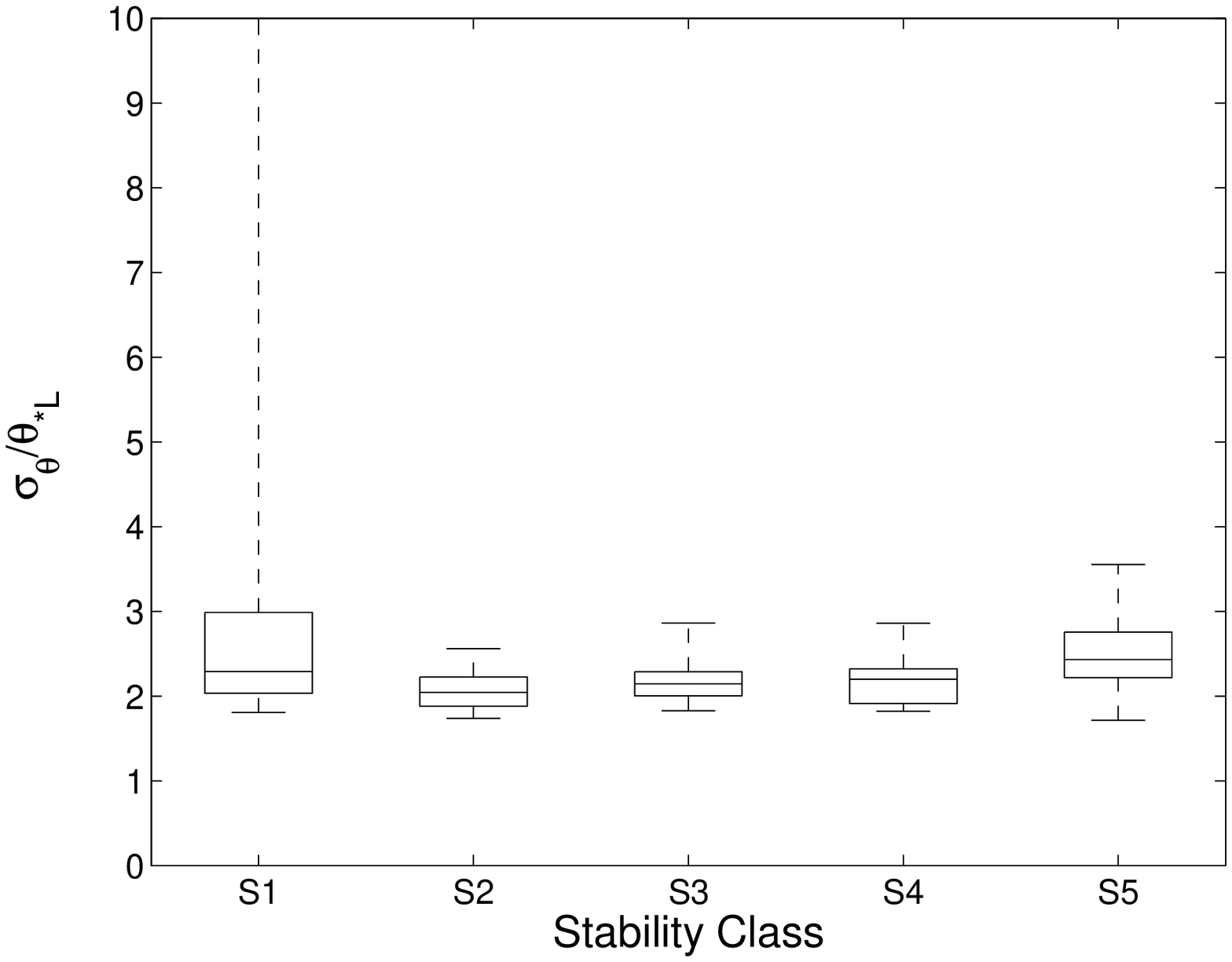}}
\centerline{\includegraphics[width=2.5in]{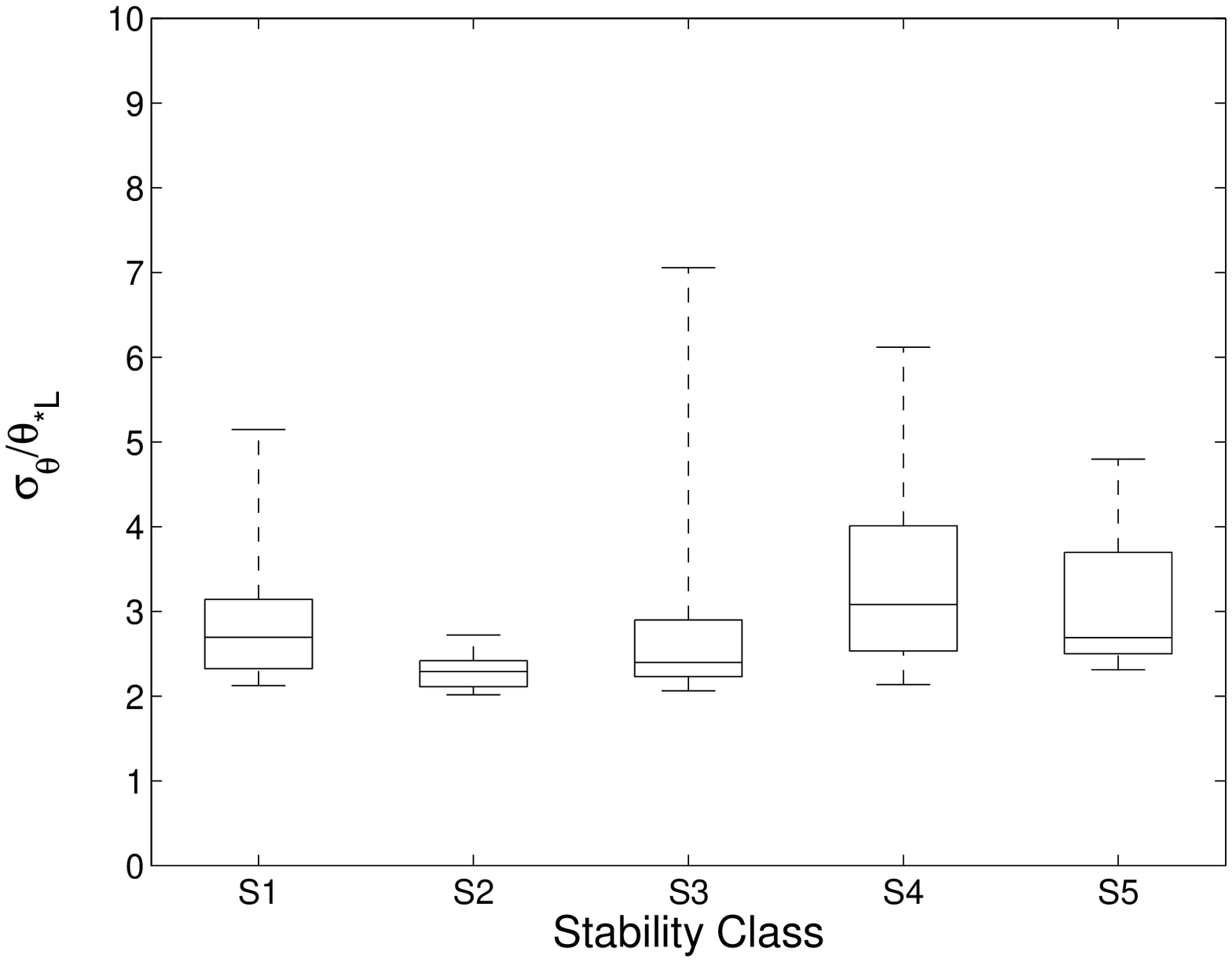}}
\centerline{\includegraphics[width=2.5in]{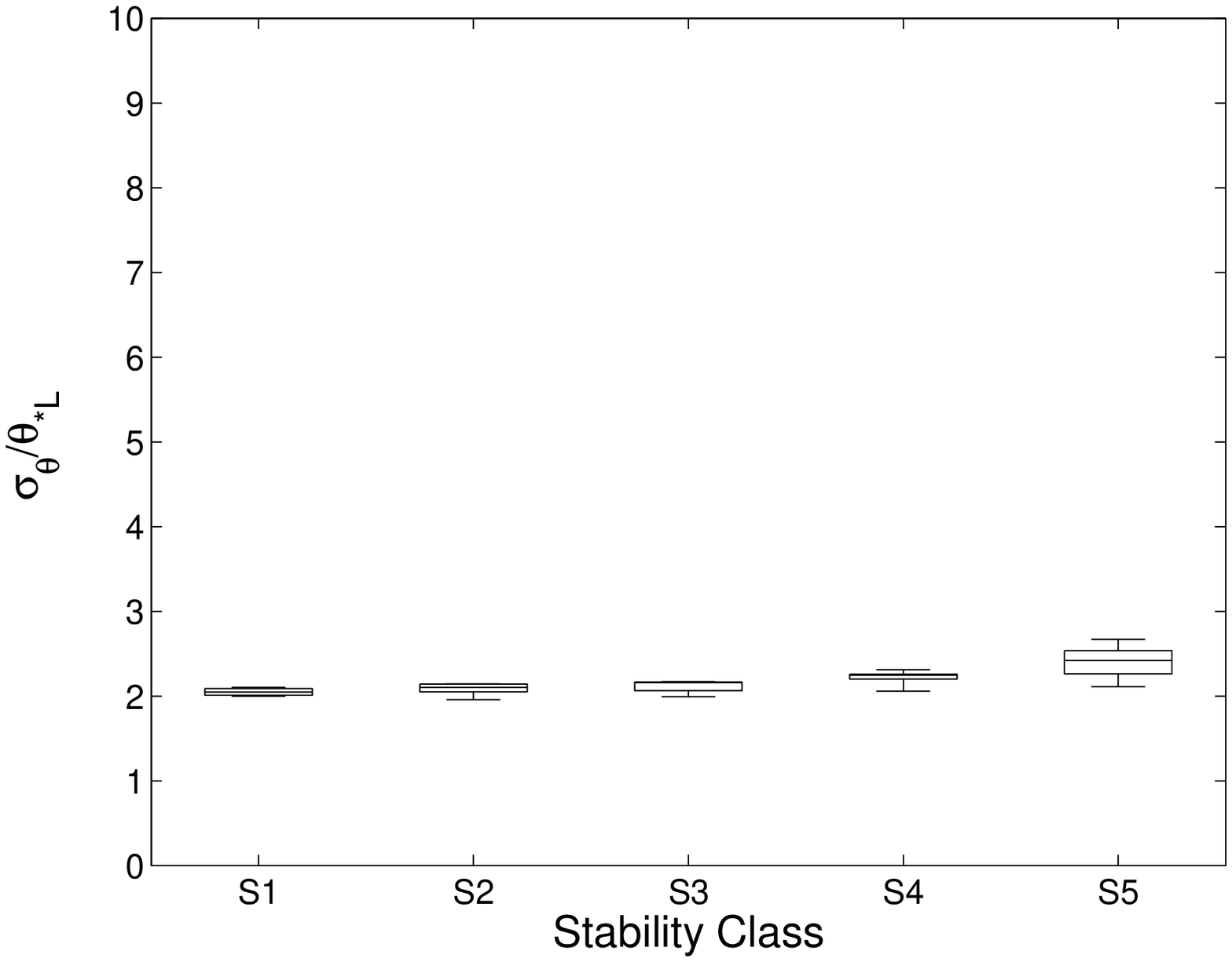}}
\caption{$\sigma_{\theta}/\theta_{*L}$  from  (top)  field  measurements,
(middle)    wind-tunnel   measurements,   and    (bottom)   large-eddy
simulations.}
\label{SigT}
\end{figure}

In Figures  \ref{SigU}, \ref{SigV}, \ref{SigW} and  \ref{SigT} we plot
the normalized standard deviation of turbulent variables.  The results
are presented  using standard boxplot  notation with marks at  95, 75,
50, 25,  and 5 percentile  of an empirical distribution.   Please note
that the Figures \ref{Fig2}  (top-right) and \ref{SigU} represent the same
results in two different formats.

It is  quite evident  from Figures \ref{SigU}  to \ref{SigT}  that the
normalized  standard  deviation of  the  turbulence variables  closely
follows    the   local   scaling    predictions   and    also   z-less
stratification. In Table \ref{T3}  we further report the median values
of   the   turbulence  statistics   corresponding   to  the   category
S5. Loosely, these median values could be considered as the asymptotic
z-less values, which are found  to be remarkably close to Nieuwstadt's
analytical  predictions and  also  his field  observations (see  Table
\ref{T3}).   For an  example,  Nieuwstadt's theory  predicts that  the
normalized   vertical  velocity   standard   deviation  asymptotically
approaches $\sim1.4$ in  the z-less regime.  In the  present study, we
observe this value to be in the narrow range of 1.4 to 1.6.  Recently,
\inlinecite{Heinemann04} compiled a list  (see Table 2 of their paper)
of  turbulence statistics under  very stable  conditions ($\zeta_{max}
\sim$ 25)  reported by various researchers.  They  found an asymptotic
value  of $\sim1.6$  for $\sigma_w/u_{*L}$.   These results  should be
contrasted with Figure 3 of \inlinecite{Pahlow}.

\begin{figure}
\centerline{\includegraphics[width=2.5in]{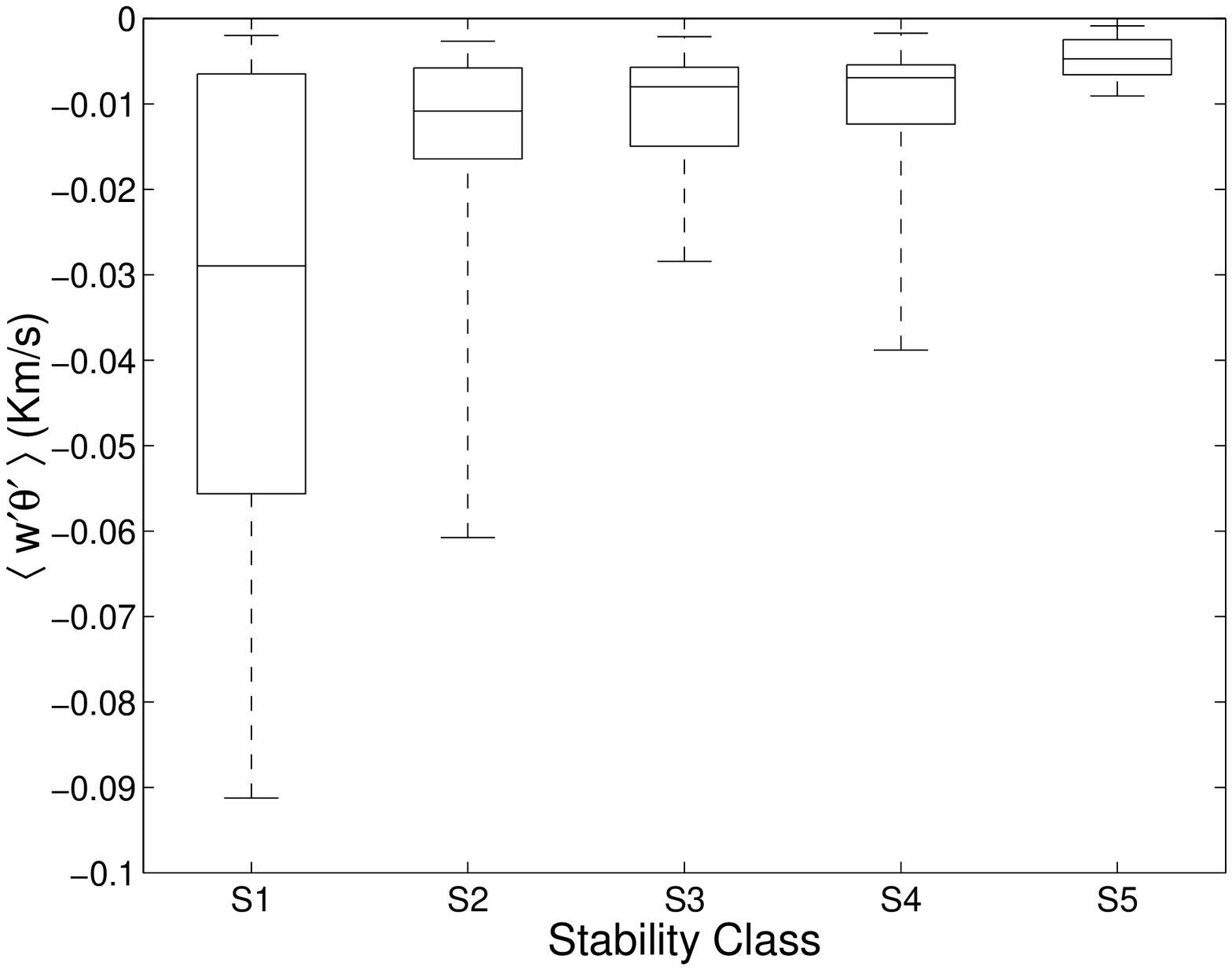}}
\centerline{\includegraphics[width=2.5in]{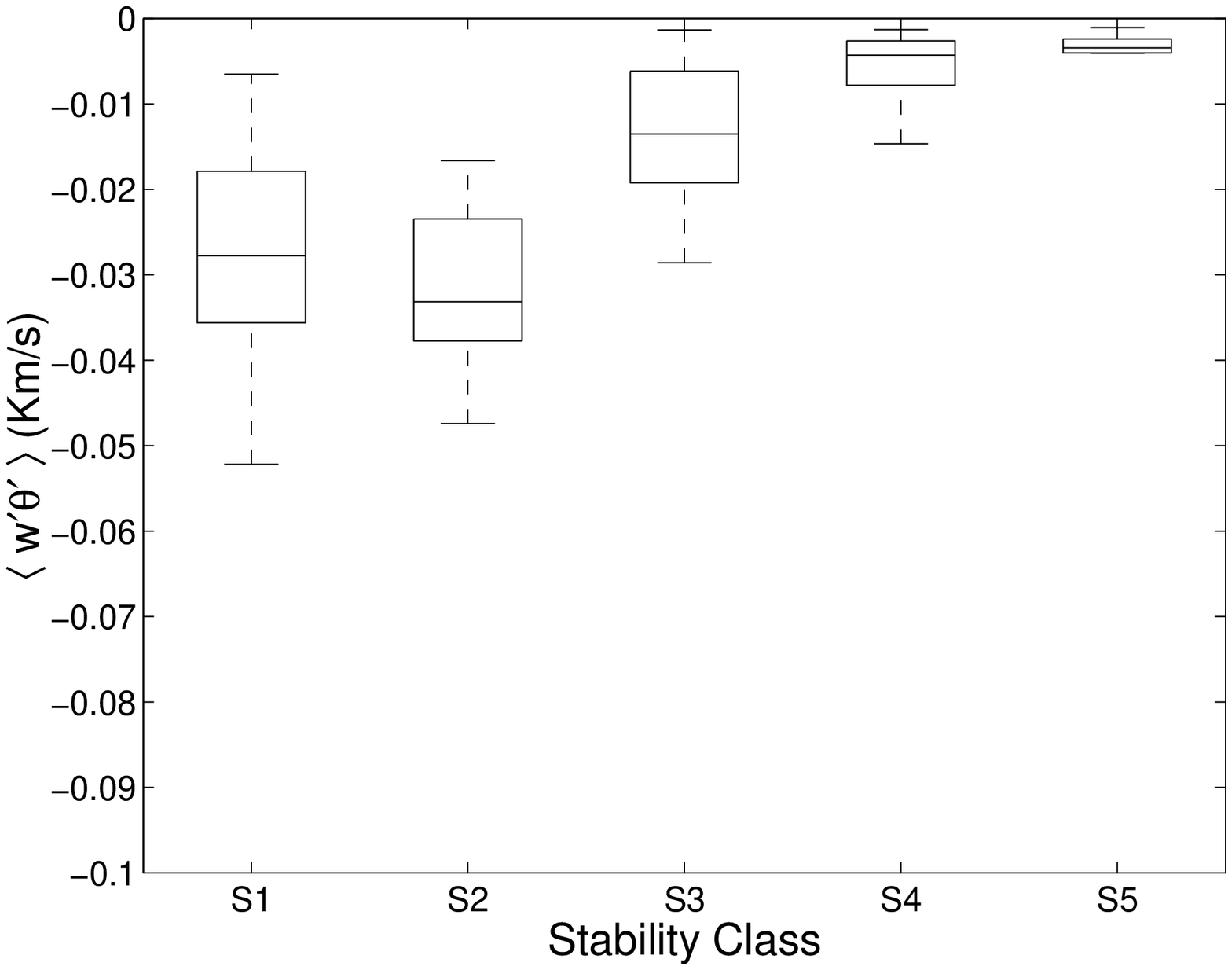}}
\centerline{\includegraphics[width=2.5in]{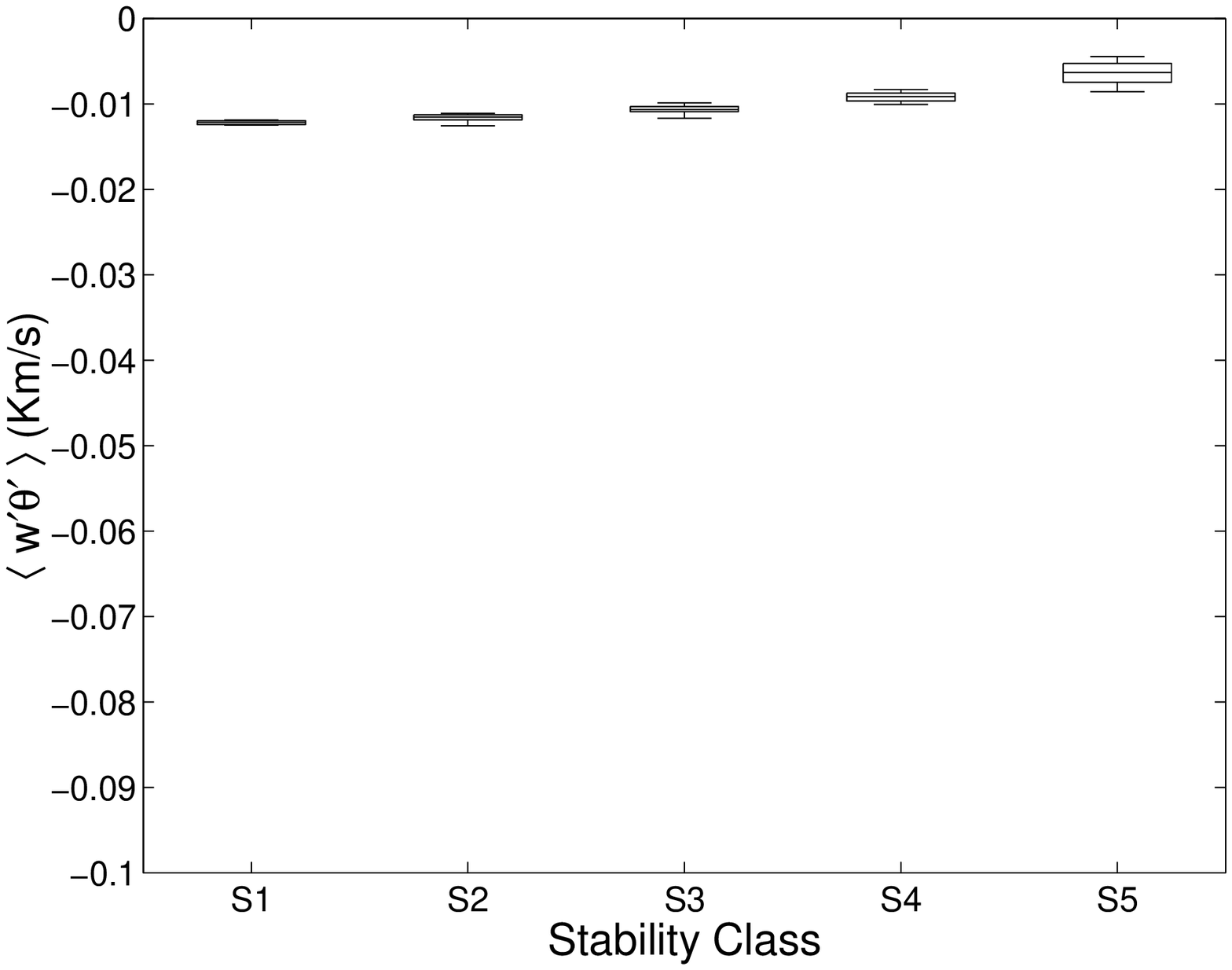}}
\caption{Heat  flux  ($\overline{w\theta}$)  (m K s$^{-1}$)  from  (top)
field  measurements, (middle)  wind-tunnel measurements,  and (bottom)
large-eddy simulations.}
\label{wT}
\end{figure}

Next, we plot the downward  heat flux profiles in Figure \ref{wT}.  In
the very  stable regime (class  S5) due to suppression  of turbulence,
the  heat flux  vanishes  \cite{Mahrt98}.  Of  course,  the heat  flux
should also go to zero in  the near-neutral limit (class S1) since the
temperature  fluctuations become  quite small.   The  maximum downward
heat flux occurs in  between these two extremes.  \inlinecite{Mahrt98}
reported  that this maximum  flux occurs  at $\zeta  = 0.05$  based on
Microfronts data, whereas \inlinecite{Mahli} found $\zeta$ to be 0.20.
In the  literature, there  is no general  consensus on this  value and
also  from Figure  \ref{wT} it  is quite  difficult to  estimate.  The
wind-tunnel measurements  show that the  maximum heat flux  happens in
the  stablity class  S2 (i.e.,  $\zeta$ =  0.10 -  0.25),  which would
support  Mahli's  result.    However,  the  field  measurements  would
definitely be in favor of \inlinecite{Mahrt98}.

\begin{figure}
\centerline{
\begin{tabular}{c@{\hspace{0.3pc}}c}
\includegraphics[width=2.2in]{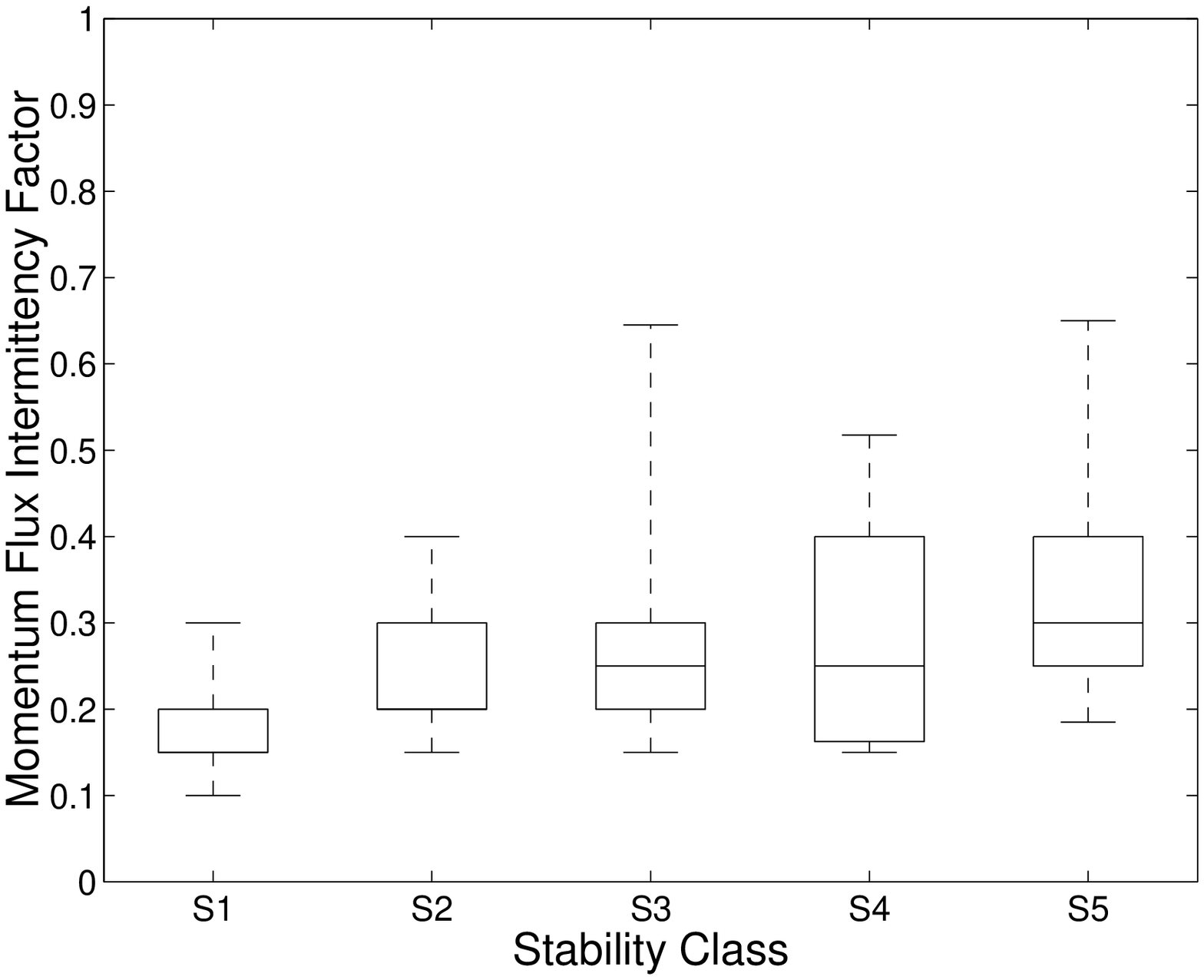}&
\includegraphics[width=2.2in]{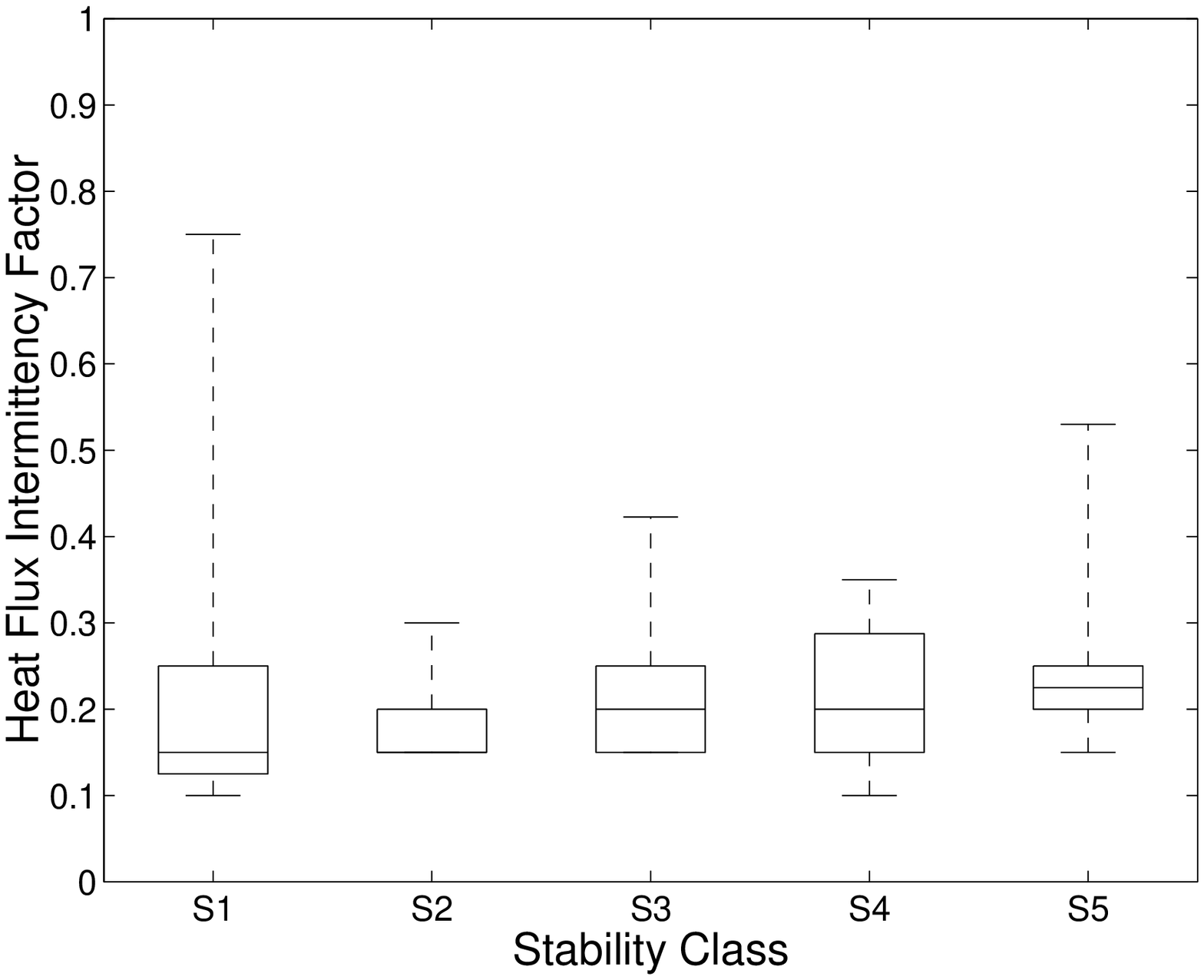}\\
\caption{Intermittency   factors  corresponding  to   (left)  momentum
fluxes, and (right) heat fluxes derived from field measurements.}
\label{IF}
\end{tabular}}
\end{figure}

It is  widely accepted that  as the stability increases  the turbulent
fluxes become  more and more intermittent \cite{Mahrt89}.   One way to
quantify  the degree of  flux intermittency  is the  use of  so called
Intermittency  Factor ($IF$),  introduced by  \inlinecite{Howell}.  To
compute $IF$, first of all  one needs to divide individual time series
of $u,~v,~w$ and $\theta$ into $N$ smaller subrecords (in this work $N
= 20$, which  correspond to 1.5 min windows for 30  min signals and so
on).   Subsequently,  local  fluxes  ($F_i$)  are  computed  from  the
deviation of  subrecord averages.  If  $M$ subrecords are  needed such
that  the ratio of  $\sum\limits_{i=1}^M F_i$  to $\sum\limits_{i=1}^N
F_i$ exceeds  $0.9$, then the  intermittency factor is  simply defined
as: $IF  = 1  - M/N$. In  the asymptotic  limit of $M  \rightarrow N$,
i.e., when  the turbulent fluxes are uniformly  distributed, $IF$ goes
to  zero.   On  the  other  hand,  if  $M  \rightarrow  1$,  then  the
intermittency factor approaches unity.  From the field measurements we
compute    the   intermittency    factors   for    vertical   momentum
($\overline{uw}$) and  heat fluxes ($\overline{w\theta}$)  (see Figure
\ref{IF}).  For the entire stability  range the $IF$s are greater than
zero.   For the  momentum  flux, the  increase  of intermittency  with
increasing stability  is quite  clear.  In the  case of heat  flux the
intermittency increases in both  the near-neutral (S1) and very stable
(S5) regimes.  In the near-neutral regime the temperature fluctuations
are  very small and  the computation  of very  weak heat  flux becomes
problematic  and leads  to significant  intermittency.  \footnote{This
intermittent near-neutral  behavior is also reflected in  the plots of
variance and third-order moment of temperature (see Figures \ref{SigT}
and \ref{Phi} respectively), as would be anticipated.} The increase in
IF in  the case of S5  is definitely a signature  of intermittent very
stable boundary layers.

\begin{figure}
\centerline{\includegraphics[width=2.5in]{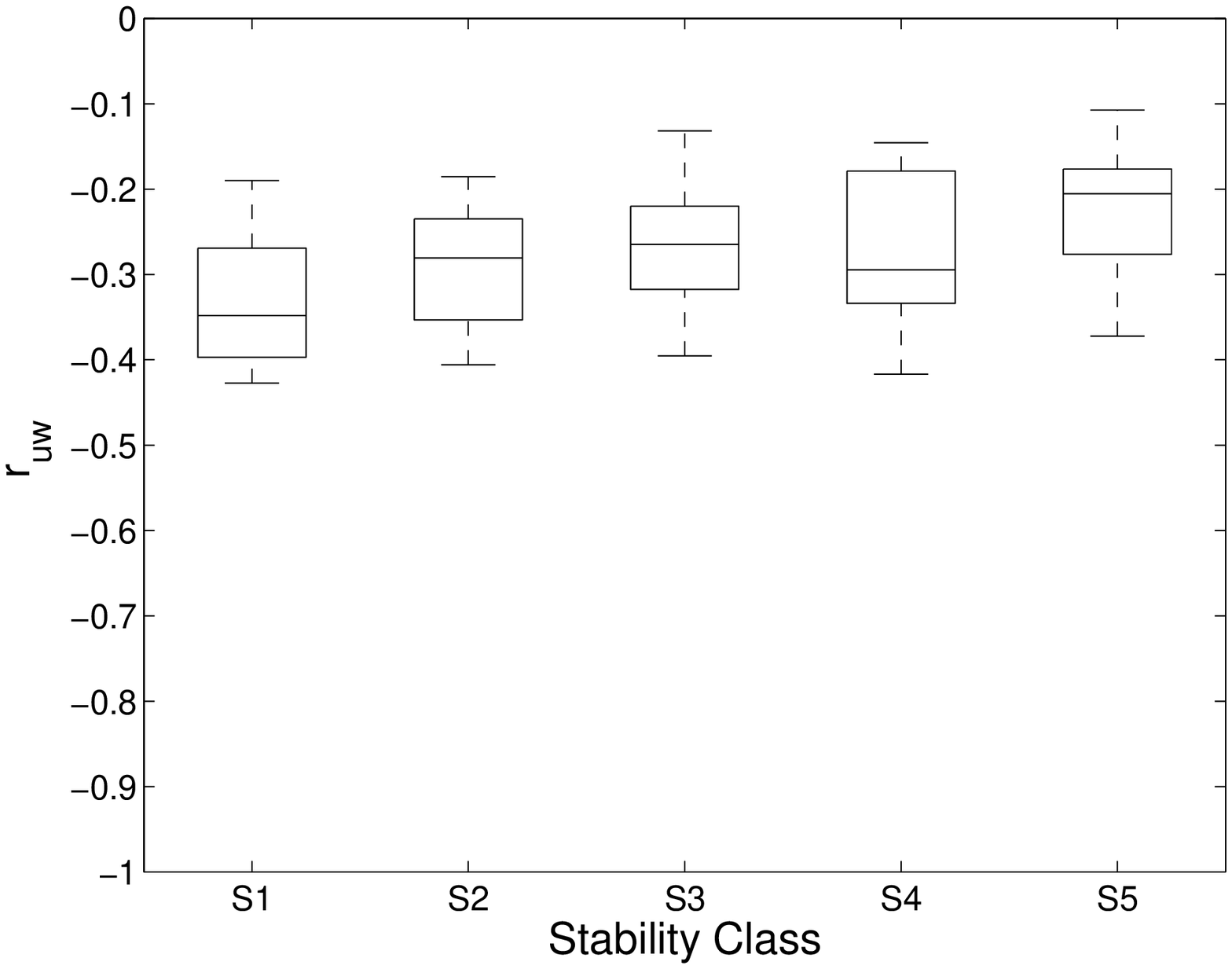}}
\centerline{\includegraphics[width=2.5in]{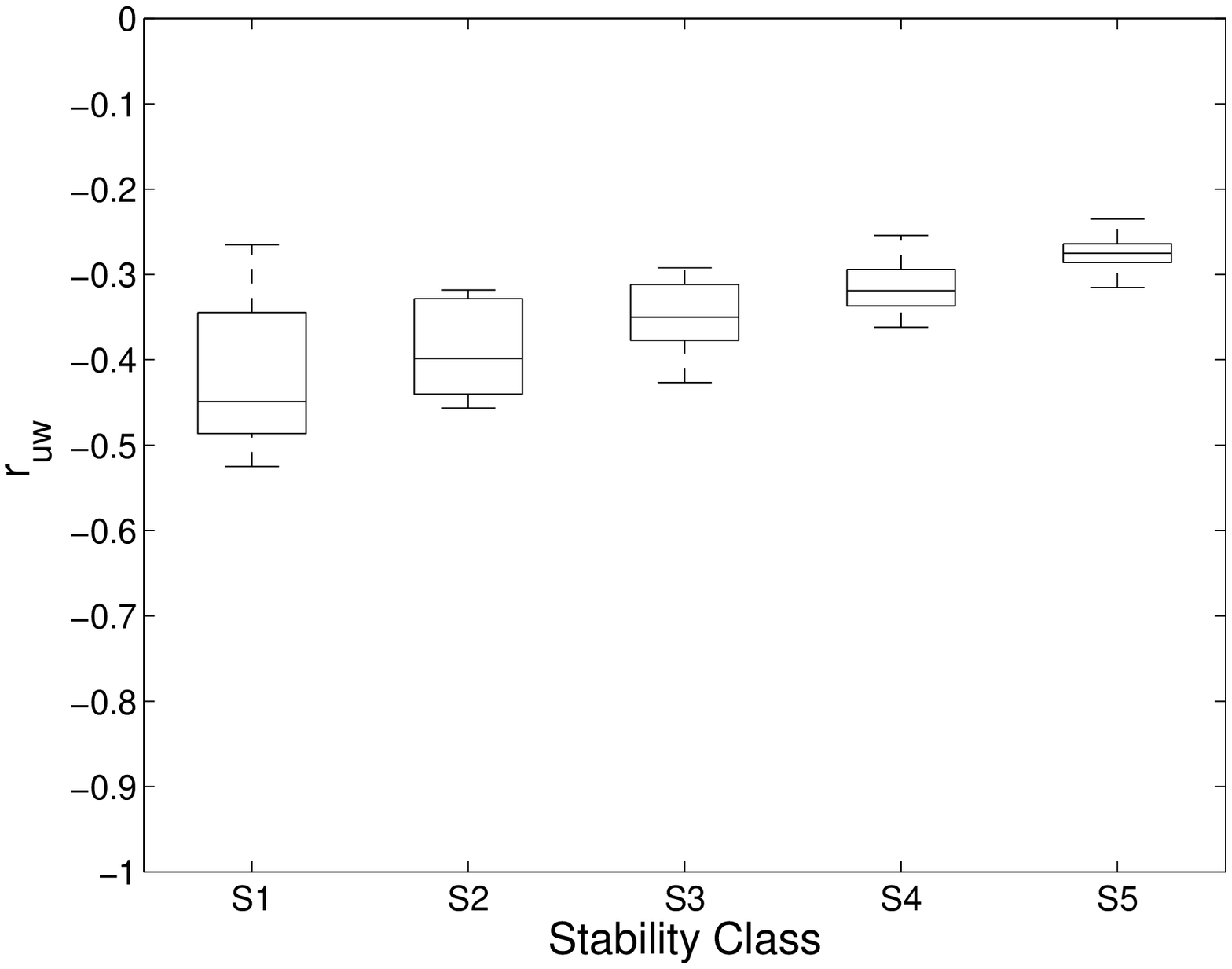}}
\centerline{\includegraphics[width=2.5in]{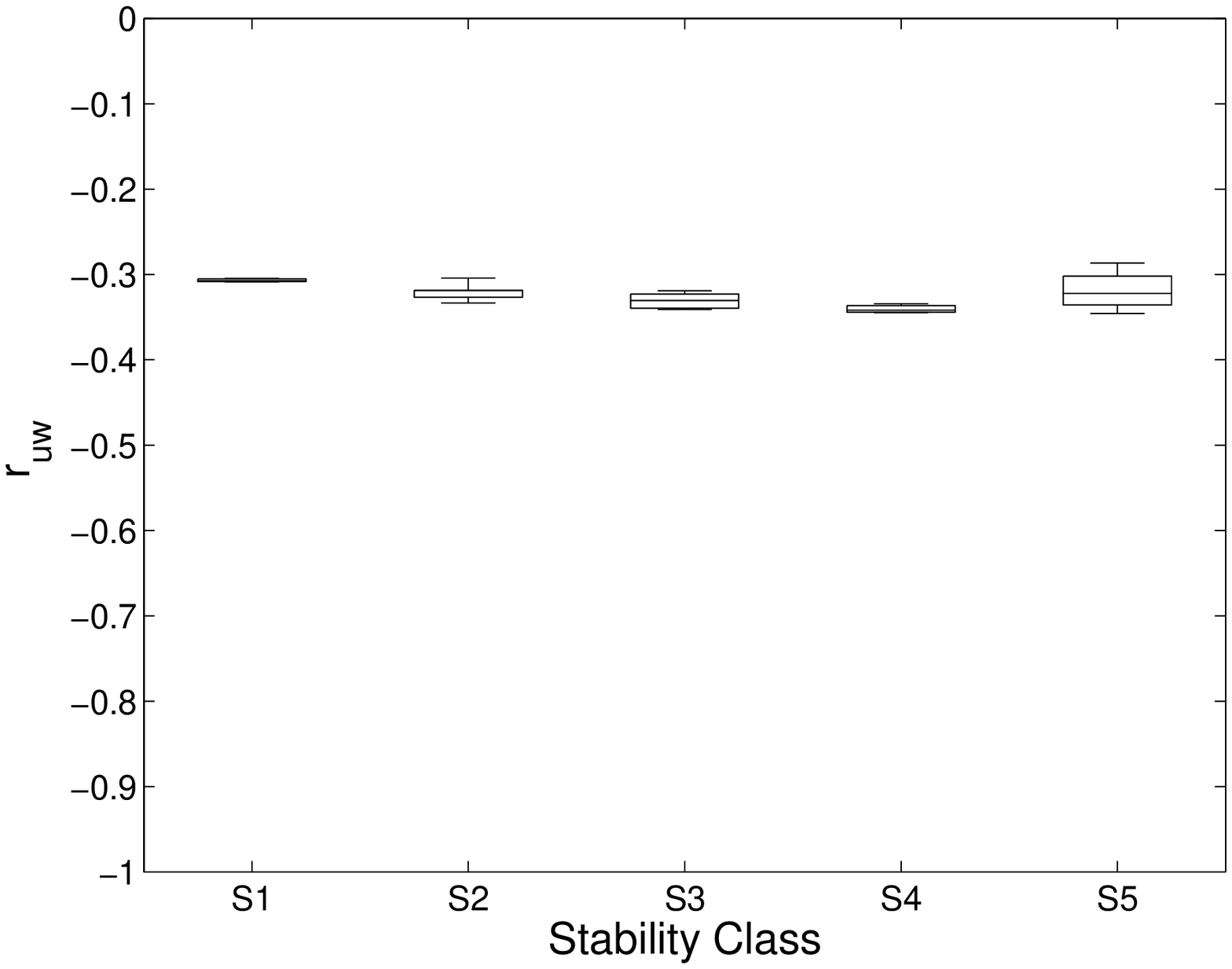}}
\caption{Correlation between  $u$ and $w$ ($r_{uw}$)  from (top) field
measurements,   (middle)   wind-tunnel   measurements,  and   (bottom)
large-eddy simulations.}
\label{ruw}
\end{figure}

\begin{figure}
\centerline{\includegraphics[width=2.5in]{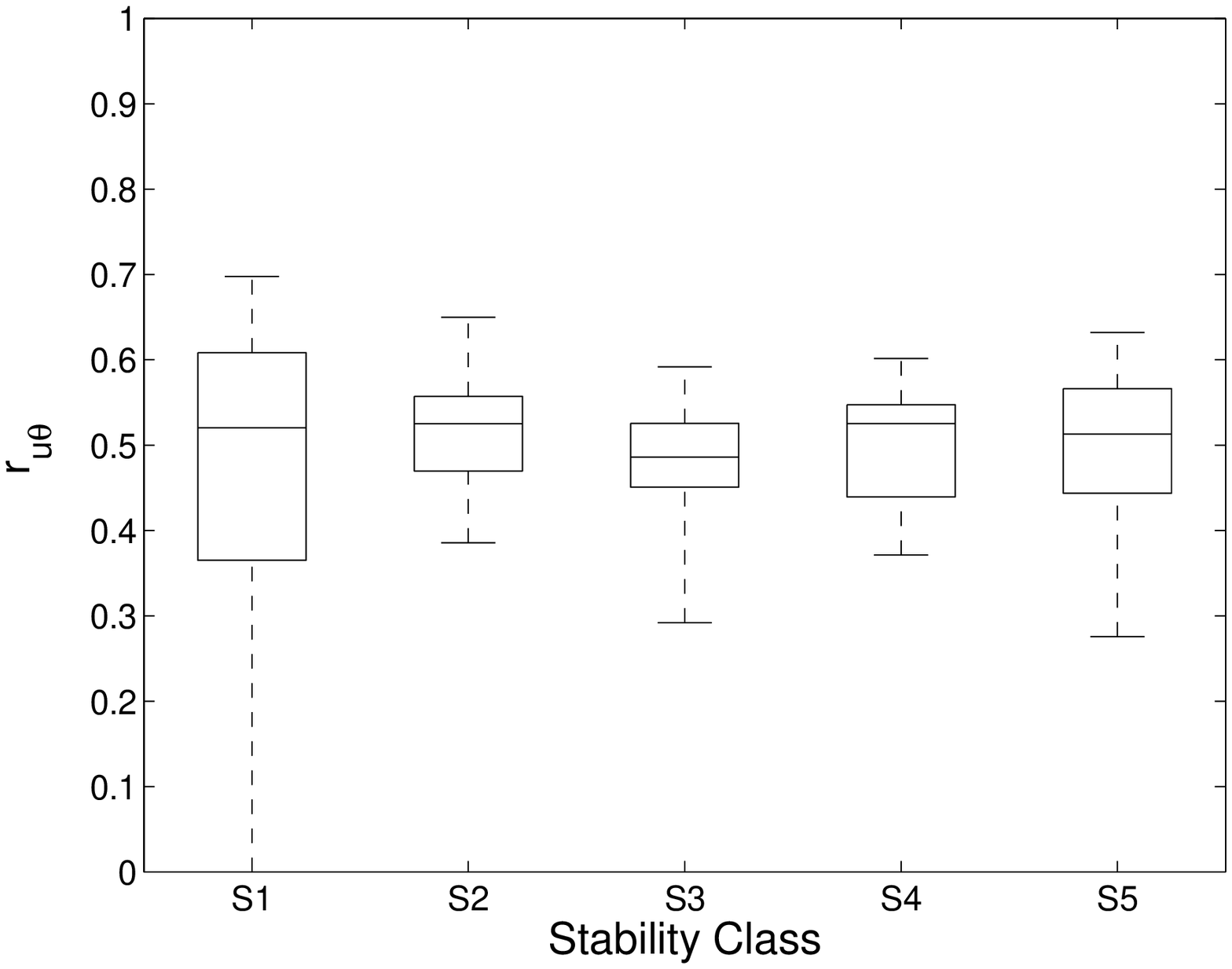}}
\centerline{\includegraphics[width=2.5in]{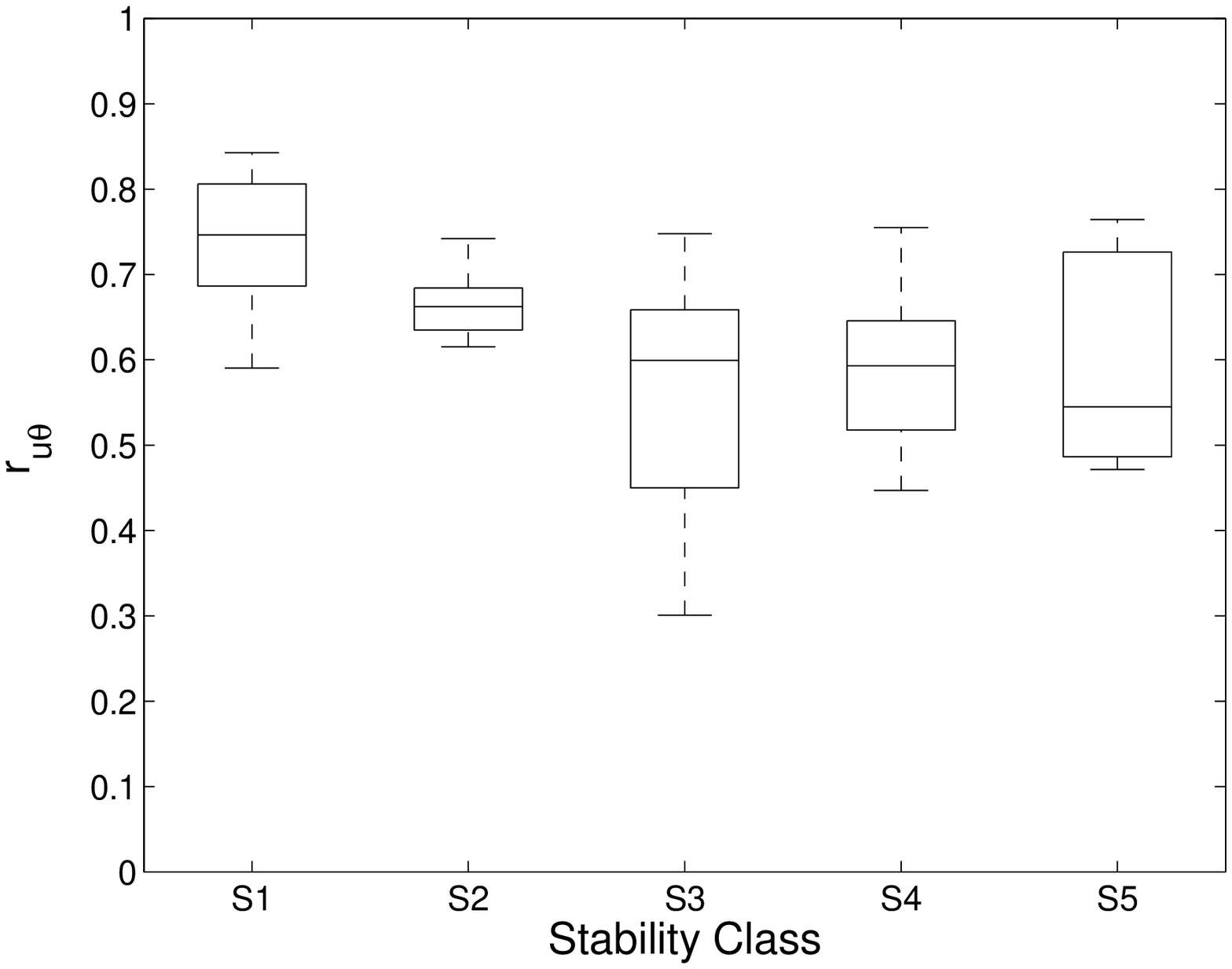}}
\centerline{\includegraphics[width=2.5in]{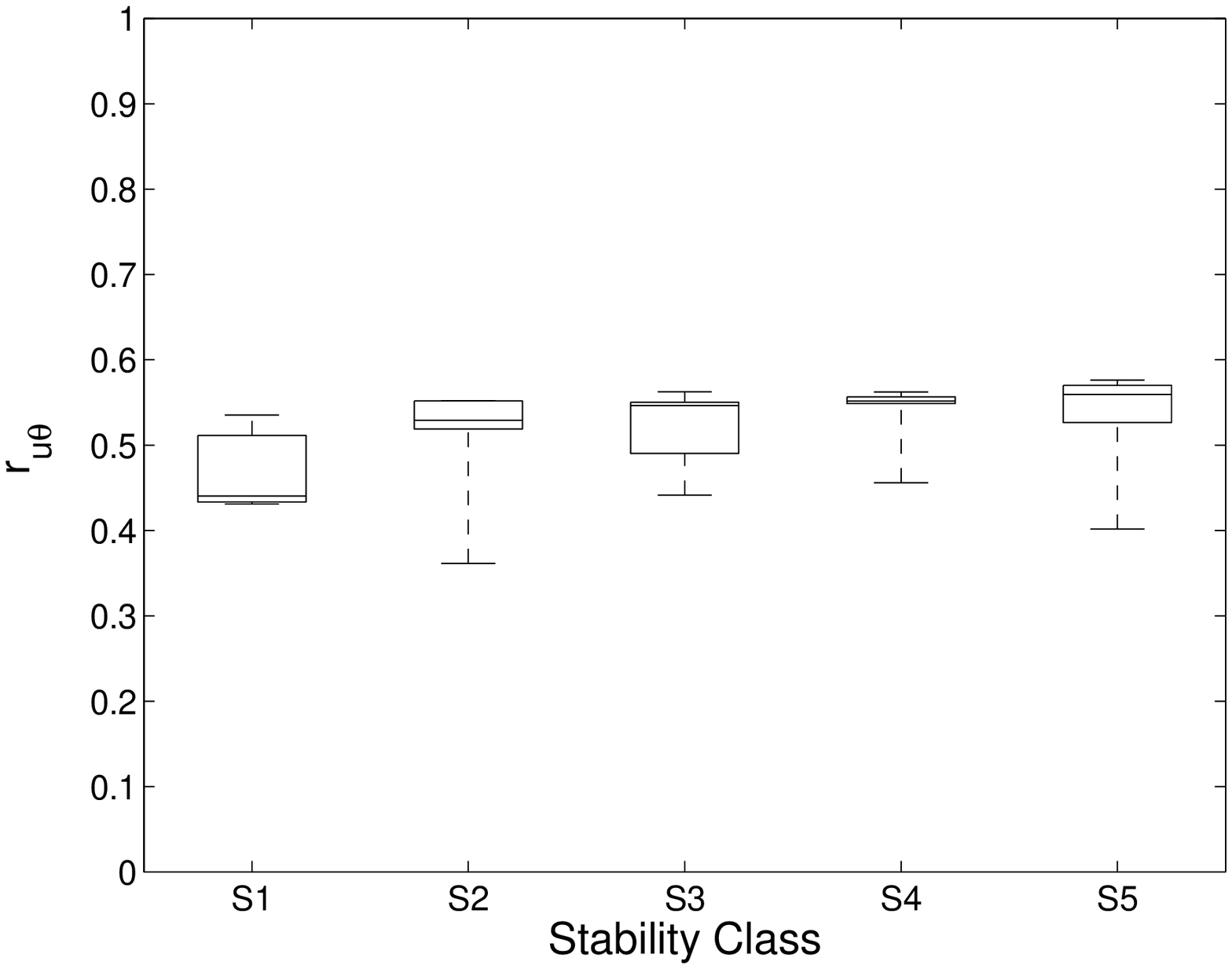}}
\caption{Correlation  between $u$  and  $\theta$ ($r_{u\theta}$)  from
(top)  field  measurements,  (middle)  wind-tunnel  measurements,  and
(bottom) large-eddy simulations.}
\label{ruT}
\end{figure}

\begin{figure}
\centerline{\includegraphics[width=2.5in]{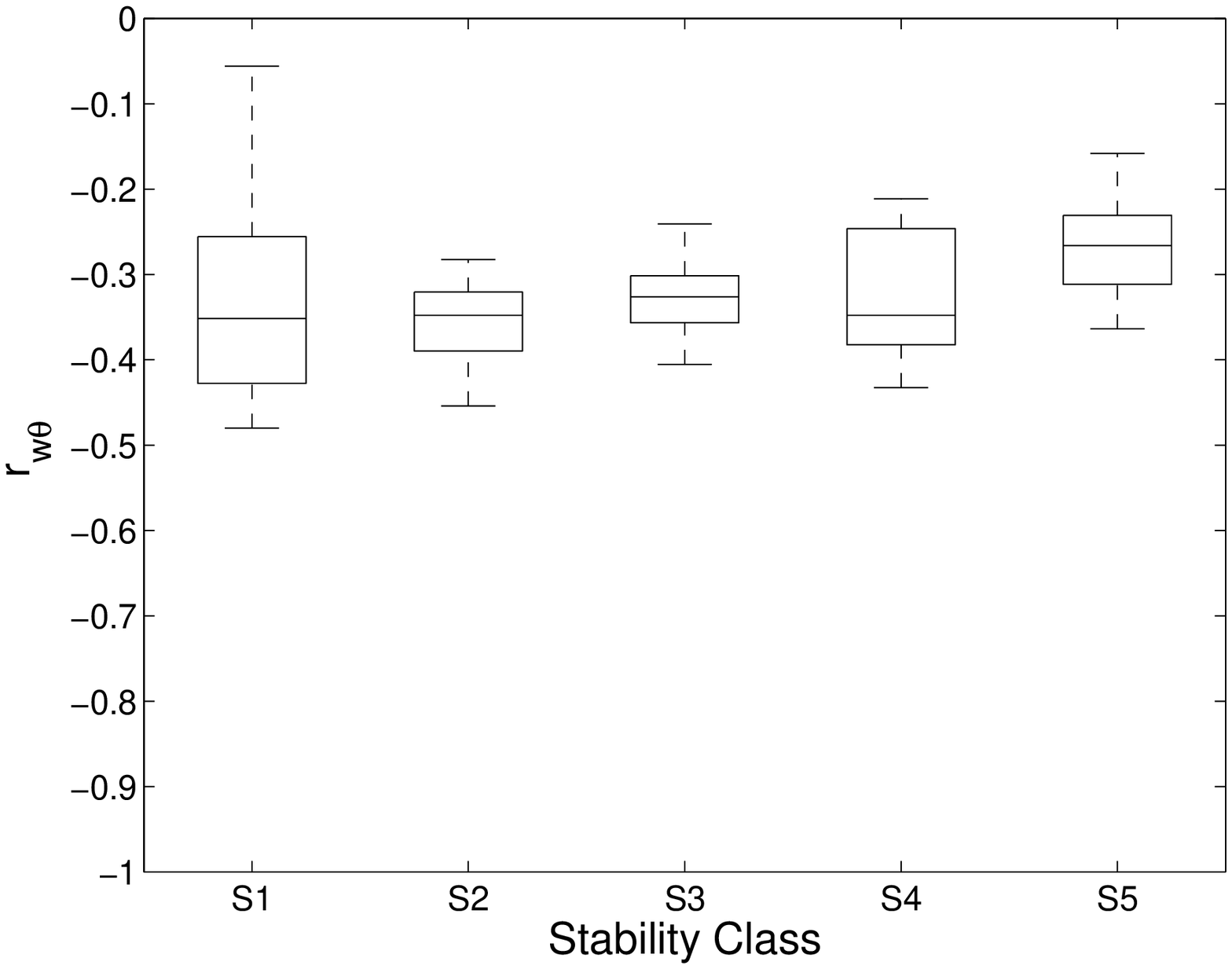}}
\centerline{\includegraphics[width=2.5in]{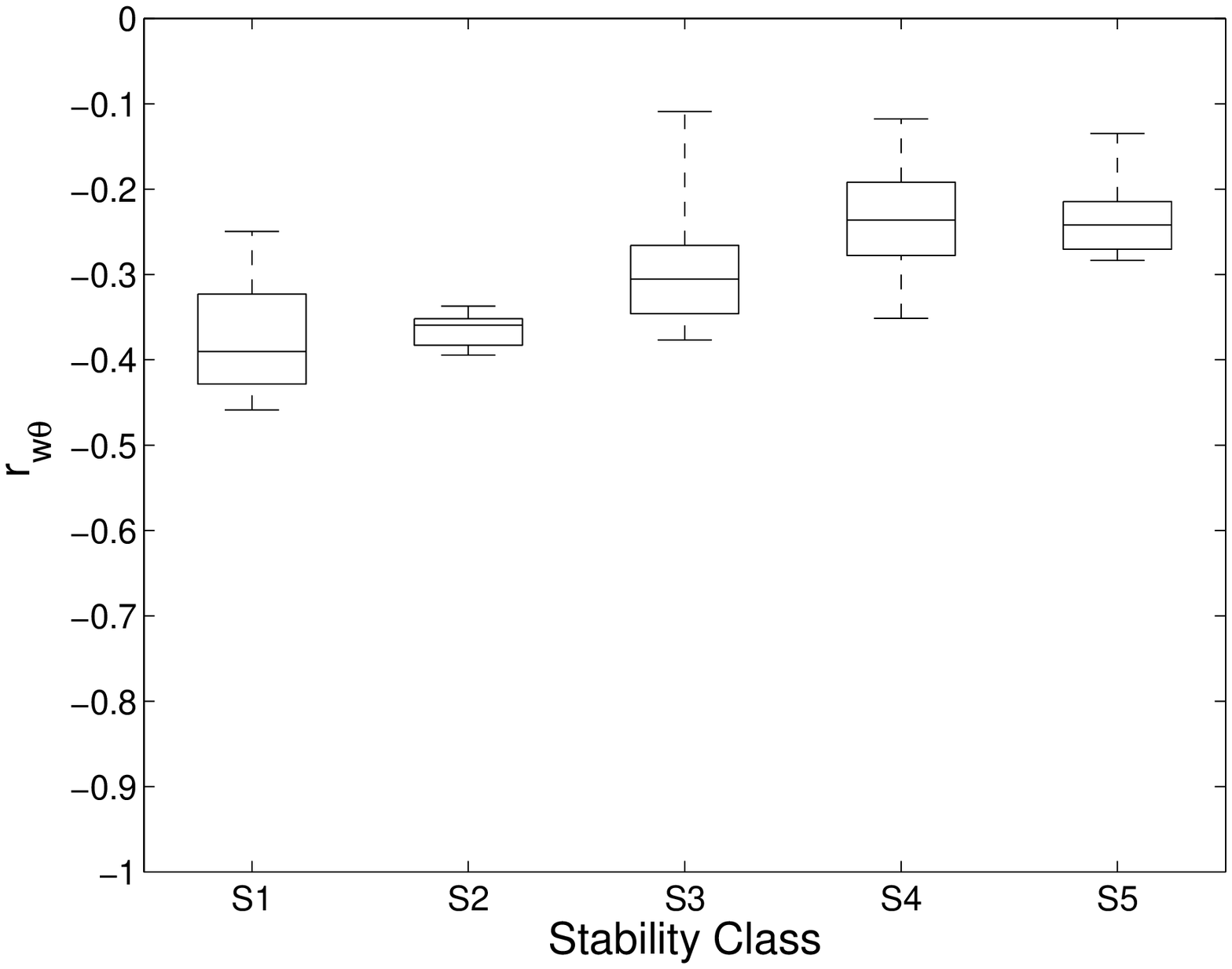}}
\centerline{\includegraphics[width=2.5in]{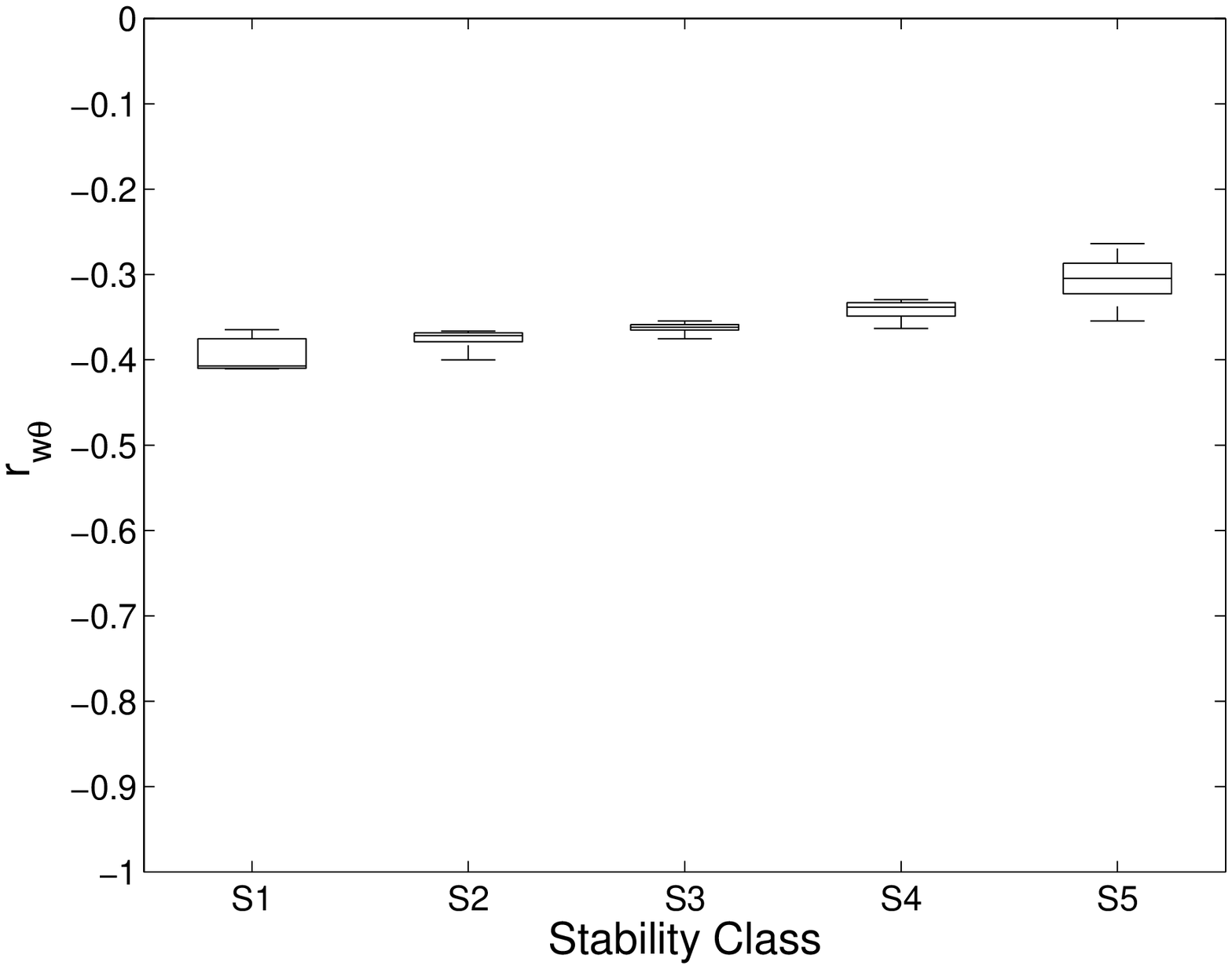}}
\caption{Correlation  between $w$  and  $\theta$ ($r_{w\theta}$)  from
(top)  field  measurements,  (middle)  wind-tunnel  measurements,  and
(bottom) large-eddy simulations.}
\label{rwT}
\end{figure}

\begin{figure}
\centerline{\includegraphics[width=2.2in]{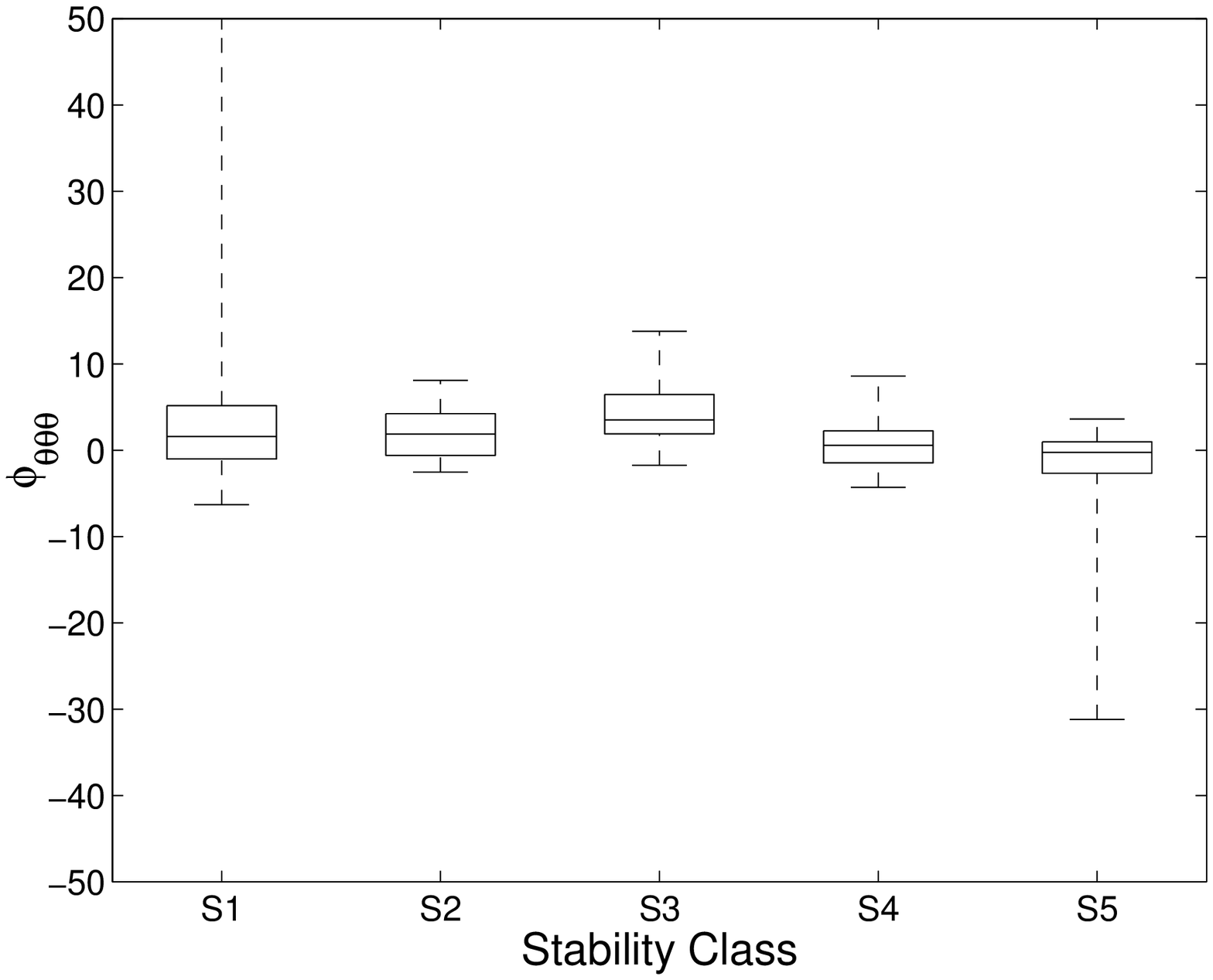}}
\centerline{\includegraphics[width=2.2in]{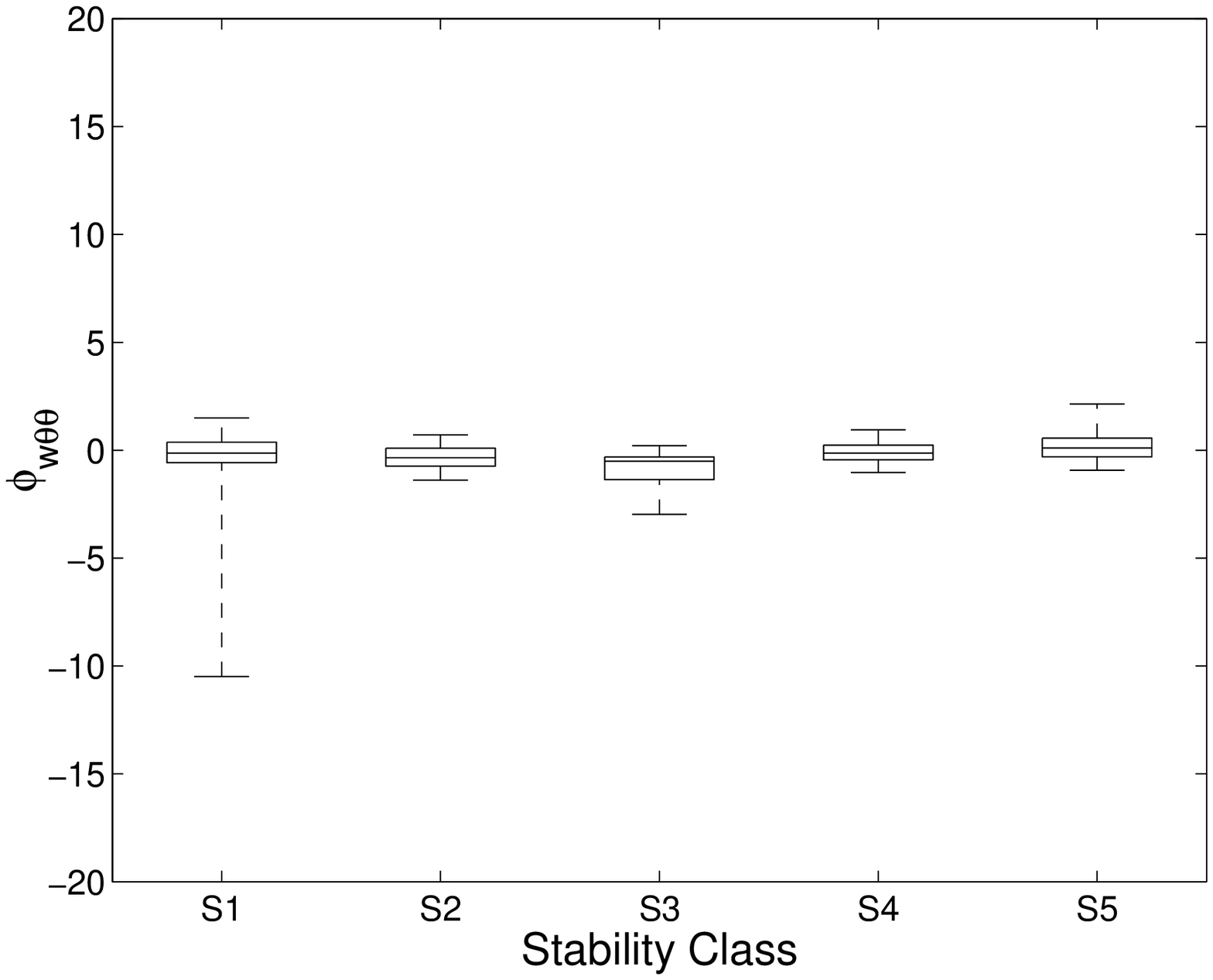}}
\centerline{\includegraphics[width=2.2in]{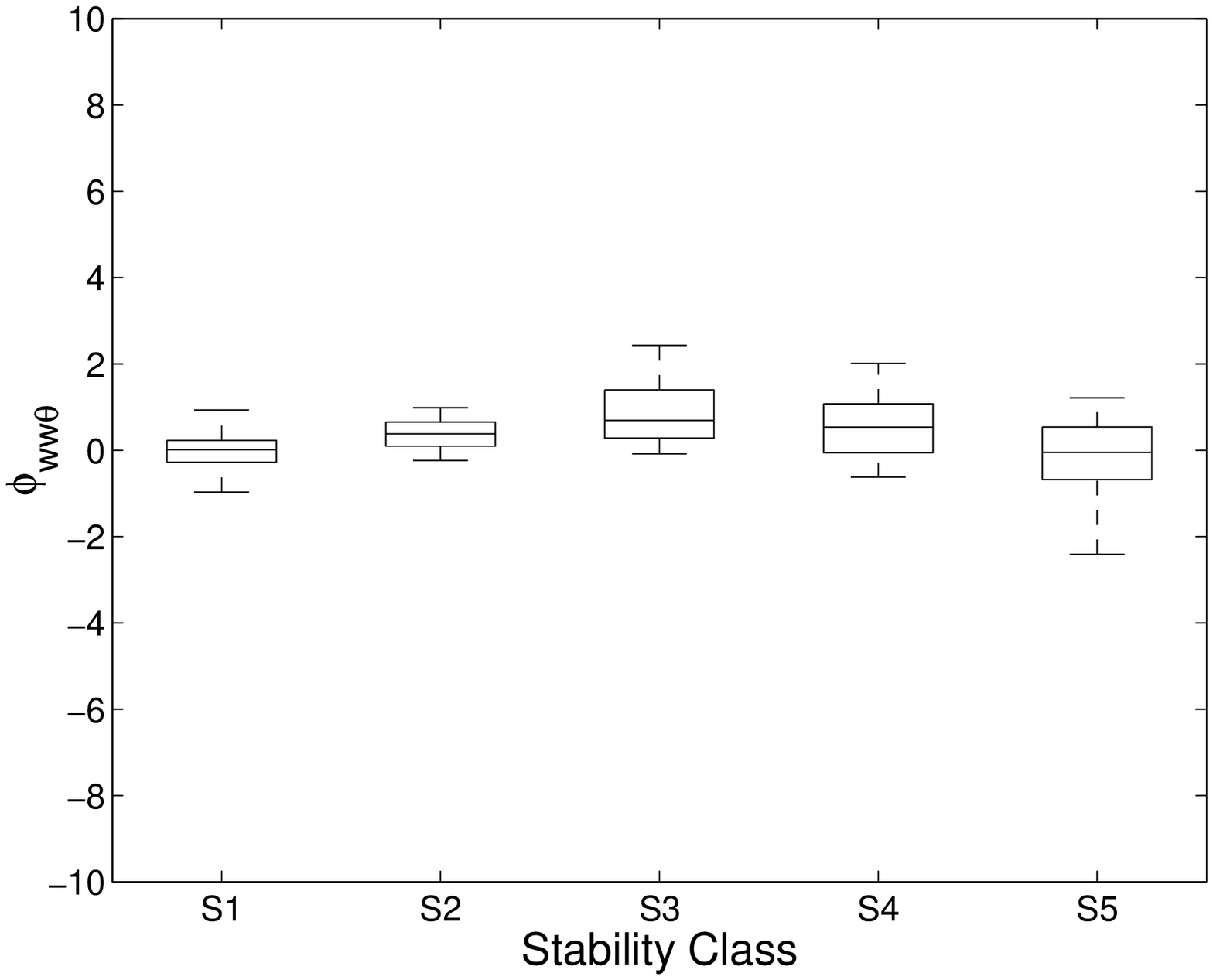}}
\caption{Nondimensionalized       third-order      moments:      (top)
$\phi_{\theta\theta\theta}$,   (middle)   $\phi_{w\theta\theta}$,  and
(bottom) $\phi_{ww\theta}$, obtained from field observations.}
\label{Phi}
\end{figure}

In Figures  \ref{ruw}, \ref{ruT} and  \ref{rwT}, we report  the mutual
correlations between  u, w and  $\theta$.  The z-less values  are also
reported in Table \ref{T3}. Once  again, these values are very similar
to  the  ones  compiled  by \inlinecite{Heinemann04}  and  theoretical
predictions    of     \inlinecite{Nieuwstadt84b}.     As    a    note,
\inlinecite{Kaimal}   also  report  that   for  $0   <  \zeta   <  1$,
$r_{u\theta}$ = 0.6, which is close to the values found in the present
study (see Figure \ref{ruT}).

\begin{table}
\caption{Median z-less values of turbulence statistics}\label{T3}
\begin{tabular}{ccccc} \hline
Turbulence &  Field & Wind  Tunnel & Large-Eddy &  Nieuwstadt \\
Statistics &  Observations  &  Measurements  & Simulations  &
(1984b, 1985)\\ \hline $\sigma_u/u_{*L}$  & 2.7 &  2.5 &
2.3   &  2.0\\  $\sigma_v/u_{*L}$   &  2.1   &  --   &  1.7   &  1.7\\
$\sigma_w/u_{*L}$    &     1.6    &     1.5    &    1.4     &    1.4\\
$\sigma_\theta/\theta_{*L}$ & 2.4 & 2.7 & 2.4 & 3.0\\ $r_{uw}$ & -0.21
&  -0.28 &  -0.32  &  -\\ $r_{u\theta}$&  0.51  & 0.55  &  0.56 &  -\\
$r_{w\theta}$& -0.27 & -0.24 & -0.30 & -0.24\\ \hline
\end{tabular} 
\end{table}

Lastly, in  Figure \ref{Phi} we  plot the stability dependence  of the
nondimensionalized  third-order moments  ($\phi_{\theta\theta\theta} =
\overline{\theta^3}/\theta_*^3$,        $\phi_{w\theta\theta}        =
\overline{w\theta^2}/(u_*\theta_*^2)$,    and    $\phi_{ww\theta}    =
\overline{ww\theta}/(u_*^2\theta_*)$)    derived   from    our   field
measurements database.   Even though in the past  several studies have
provided  evidence  of   local-scaling  in  turbulence  gradients  and
variances,   results  confirming   its  existence   in  the   case  of
higher-order  moments are  quite rare  in the  literature.   A notable
exception was the study  by \inlinecite{Dias}.  They showed that these
nondimensionalized   third-order  moments   obey  local   scaling  and
essentially remain constant ($\sim$  0) for the entire stability range
considered.  As  evident from  Figure \ref{Phi}, our  present analysis
definitely supports the conclusions of \inlinecite{Dias}.

In light of  the foregoing analyses and discussion  it is certain that
the  local scaling hypothesis  of Nieuwstadt,  which has  survived the
last two decades, still holds for a wide range of stabilities provided
that mesoscale motions are not included.

\section{Summary}\label{Sec6}

In  this study, we  performed rigorous  statistical analyses  of field
observations  and   wind-tunnel  measurements  and   also  employed  a
new-generation large-eddy SGS model in order to verify the validity of
Nieuwstadt's  local-scaling hypothesis  under very  stable conditions.
An  extensive set of  turbulence statistics,  computed from  field and
wind-tunnel measurements or from  LES generated datasets, supports the
validity of the local scaling  hypothesis (in the cases of traditional
bottom-up  as   well  as  upside-down  stable   boundary  layers  over
homogeneous,  flat  terrains).    We  demonstrate  that  non-turbulent
effects need to  be removed from field data  while studying similarity
hypotheses, otherwise the results could be misleading.

In  a parallel  work \cite{Basu},  we  also found  that the  stability
functions  (commonly  used  in  the  first-order  turbulent  K-closure
models)   extracted   from  idealized   LESs   closely  resemble   the
field-observations-based  M-O stability functions  \cite{Basu}.  These
kinds  of   agreements  between   our  simulated  results   and  field
observations  are  very  encouraging.   They  not  only  provide  more
confidence in our results but  also highlight the credibility  of our
scale-dependent  dynamic SGS  modeling approach  in  simulating stable
boundary layers.

\acknowledgements Special  thanks go to  Yuji Ohya for sending  us his
state-of-the-art  wind-tunnel  data.  We  are  grateful  to all  those
researchers who  painstakingly collected  data during the  Iowa, Davis
and  CASES-99 field  campaigns.  We  greatly acknowledge  the valuable
comments and suggestions  made by Rob Stoll during  the course of this
study.  This work was partially funded by NSF and NASA grants.  One of
us  (SB)   was  partially  supported  by   the  Doctoral  Dissertation
Fellowship  from the  University of  Minnesota. All  the computational
resources  were  kindly   provided  by  the  Minnesota  Supercomputing
Institute.

\end{article}

\begin{thebibliography}{}

\bibitem[\protect\citeauthoryear{Albertson                          and
Parlange}{1999}]{Albertson}  Albertson,  J.~D.,  and Parlange,  M.~B.:
1999, \newblock  {`Natural Integration  of Scalar Fluxes  from Complex
Terrain'}, \newblock {\em Adv. Wat. Res.}  \textbf{23}, 239--252.

\bibitem[\protect\citeauthoryear{Andr\'{e}n}{1995}]{Andren}
Andr\'{e}n, A.:  1995, \newblock {`The Structure  of Stably Stratified
Atmospheric   Boundary  Layers:   A  Large-Eddy   Simulation  Study'},
\newblock {\em Quart. J. Roy. Meteorol. Soc.} \textbf{121}, 961--985.

\bibitem[\protect\citeauthoryear{Arya}{2001}]{Arya} Arya, S. P.: 2001,
\newblock {\em Introduction  to Micrometeorology}, Academic Press, San
Diego, CA, 420 pp.

\bibitem[\protect\citeauthoryear{Barnard}{2000}]{Barnard}      Barnard,
J. C.:  2000, \newblock {`Intermittent  Turbulence in the  Very Stable
Ekman  Layer'}, \newblock  {\em PhD  Thesis, Department  of Mechanical
Engineering, University of Washington}, 154 pp.

\bibitem[\protect\citeauthoryear{Basu et al.}{2002}]{Basu02} Basu, S.,
Foufoula-Georgiou,  E.,   and  Port\'{e}-Agel,  F.:   2002,  \newblock
{`Predictability of Atmospheric Boundary-layer  Flows as a Function of
Scale'},   \newblock  {\em   Geophys.    Res.   Lett.}    \textbf{29},
doi:10.1029/2002GL015497.

\bibitem[\protect\citeauthoryear{Basu}{2004}]{Basu}      Basu,
S.:  2004, \newblock {`Large-eddy Simulation of Stably Stratified 
Atmospheric Boundary Layer Turbulence: A Scale-Dependent Dynamic Modeling
Approach'}, \newblock  {\em PhD  Thesis, Department  of Civil
Engineering, University of Minnesota}, 114 pp.


\bibitem[\protect\citeauthoryear{Beare   and   MacVean}{2004}]{Beare1}
Beare,  R.~J.,  and   MacVean,  M.~K.:  2004,  \newblock  {`Resolution
Sensitivity  and  Scaling  of  Large-Eddy Simulations  of  the  Stable
Boundary    Layer'},   \newblock   {\em    Boundary-Layer   Meteorol.}
\textbf{112}, 257--281.

\bibitem[\protect\citeauthoryear{Beare  et  al.}{2005}]{Beare2} Beare,
R.~J.,  and   Coauthors:  2004,  \newblock   {`An  Intercomparison  of
Large-Eddy Simulations of the  Stable Boundary Layer'}, \newblock {\em
Boundary-Layer Meteorol.} In Press.

\bibitem[\protect\citeauthoryear{Beljaars}{1992}]{Beljaars}   Beljaars,
A.: 1992,  \newblock {`The Parameterization of  the Planetary Boundary
Layer'}, \newblock  {\em ECMWF Meteorological  Training Course Lecture
Series}, 1--57.


\bibitem[\protect\citeauthoryear{Beljaars  and Viterbo}{1998}]{BelVit}
Beljaars,  A., and  Viterbo, P.:  1998,  \newblock {`The  Role of  the
Boundary  Layer   in  a  Numerical  Weather   Prediction  Model'},  in
A. A. M. Holtslag and P. G. Duynkerke (eds.), \newblock {\em Clear and
Cloudy  Boundary  Layers},  Royal  Netherlands  Academy  of  Arts  and
Sciences, Amsterdam, 297-304.

\bibitem[\protect\citeauthoryear{Brown    et    al.}{1994}]{Brownetal}
Brown,  A.~R., Derbyshire,  S.~H., and  Mason, P.~J.:  1994, \newblock
{`Large-Eddy Simulation  of Stable Atmospheric Boundary  Layers with a
Revised     Stochastic     Subgrid     Model'},     \newblock     {\em
Quart. J. Roy. Meteorol. Soc.} \textbf{120}, 1485--1512.


\bibitem[\protect\citeauthoryear{Canuto et al.}{1988}]{Canuto} Canuto,
C., Hussaini, M.~Y., Quarteroni, A., and Zhang, T.~A.: 1988, \newblock
{\em  Spectral Methods  in Fluid  Dynamics}, Springer  Verlag, Berlin,
Germany, 557 pp.

\bibitem[\protect\citeauthoryear{Caughey}{1982}]{Caughey}      Caughey,
S.~J.: 1982,  \newblock {`Observed Characteristics  of the Atmospheric
Boundary  Layer'}, in  F.~T.~M.  Nieuwstadt  and H.   van  Dop (eds.),
\newblock  {\em Atmospheric Turbulence  and Air  Pollution Modelling},
D. Reidel Publishing Company, Dordrecht, 107-158.



\bibitem[\protect\citeauthoryear{Cuxart   et   al.}{2000}]{Cuxartetal}
Cuxart,  J.,  and  Coauthors:  2000,  \newblock  {`Stable  Atmospheric
Boundary-Layer Experiment in Spain  (SABLES 98): A Report'}, \newblock
{\em Boundary-Layer Meteorol.} \textbf{96}, 337--370.



\bibitem[\protect\citeauthoryear{Derbyshire}{1990}]{Derby90}
Derbyshire,  S.~H.:  1990,  \newblock {`Nieuwstadt's  Stable  Boundary
Layer Revisited'},  \newblock {\em Quart.  J.   Roy.  Meteorol.  Soc.}
\textbf{116}, 127--158.

\bibitem[\protect\citeauthoryear{Derbyshire}{1999}]{Derby99}
Derbyshire, S.~H.: 1999,  \newblock {`Stable Boundary-Layer Modelling:
Established  Approaches and  Beyond'},  \newblock {\em  Boundary-Layer
Meteorol.} \textbf{90}, 423--446.

\bibitem[\protect\citeauthoryear{Dias  et  al.}{1995}]{Dias} Dias,  N.
L.,  Brutsaert,  W., and  Wesely,  M.   L.:  1995, \newblock  {`Z-Less
Stratification    under    Stable    Conditions'},   \newblock    {\em
Boundary-Layer Meteorol.} \textbf{75}, 175--187.

\bibitem[\protect\citeauthoryear{Ding  et al.}{2001}]{Ding}  Ding, F.,
Arya, S. P., and  Lin, Y.-L.: 2001, \newblock {`Large-eddy Simulations
of  the Atmospheric Boundary  Layer Using  a New  Subgrid-scale Model:
Part  II.   Weakly  and  Moderately  Stable  Cases'},  \newblock  {\em
Environ. Fluid Mech.} \textbf{1}, 49--69.


\bibitem[\protect\citeauthoryear{Forrer   and   Rotach}{1997}]{Forrer}
Forrer, J.   and Rotach, M.   W.: 1997, \newblock {`On  the Turbulence
Structure in the Stable Boundary Layer over the Greenland Ice Sheet'},
\newblock {\em Boundary-Layer Meteorol.} \textbf{85}, 111--136.

\bibitem[\protect\citeauthoryear{Galmarini  et  al.}{1998}]{Galmarini}
Galmarini, S.,  Beets, C., Duynkerke,  P.  G., and  Vil\`{a}-Guerau de
Arellano, J.:  1998, \newblock  {`Stable Nocturnal Boundary  Layers: A
Comparison  of  One-dimensional  and Large-eddy  Simulation  Models'},
\newblock {\em Boundary-Layer Meteorol.} \textbf{88}, 181--210.

\bibitem[\protect\citeauthoryear{Germano    et    al.}{1991}]{Germano}
Germano, M., Piomelli, U., Moin, P., and Cabot, W.~H.: 1991, \newblock
{`A  Dynamic  Subgrid-scale  Eddy  Viscosity Model'},  \newblock  {\em
Phys. Fluids A} \textbf{3}, 1760--1765.

\bibitem[\protect\citeauthoryear{Geurts}{2003}]{Geurts} Geurts, B.~J.:
1990, \newblock  {\em Elements  of Direct and  Large-eddy Simulation},
Edwards, Philadelphia, 329 pp.

\bibitem[\protect\citeauthoryear{Ghosal et al.}{1995}]{Ghosal} Ghosal,
S.,  Lund, T.~S.,  Moin, P.,  and Akselvoll,  K.: 1995,  \newblock {`A
Dynamic  Localization  Model for  Large-Eddy  Simulation of  Turbulent
Flows'}, \newblock {\em Phys. Fluids A} \textbf{3}, 1760--1765.

\bibitem[\protect\citeauthoryear{Heinemann}{2004}]{Heinemann04}
Heinemann, G.:  2004, \newblock  {`Local Similarity Properties  of the
Continuously  Turbulent   Stable  Boundary  Layer   over  Greenland'},
\newblock {\em Boundary-Layer Meteorol.} \textbf{112}, 283--305.

\bibitem[\protect\citeauthoryear{H\"{o}gstr\"{o}m}{1990}]{Hogstrom90}
H\"{o}gstr\"{o}m,  U.:   1990,  \newblock  {`Analysis   of  Turbulence
Structure in the Surface  Layer with a Modified Similarity Formulation
for  Near  Neutral  Conditions'},  \newblock {\em  J.   Atmos.   Sci.}
\textbf{47}, 1949--1972.

\bibitem[\protect\citeauthoryear{Holtslag}{2003}]{Holtslag}   Holtslag,
A. A. M.: 2003, \newblock {`GABLS Initiates Intercomparison for Stable
Boundary Layer Case'}, \newblock {\em GEWEX News} \textbf{13}, 7--8.

\bibitem[\protect\citeauthoryear{Howell     and    Sun}{1999}]{Howell}
Howell, J. F.  and Sun,  J.: 1999, \newblock {`Surface-Layer Fluxes in
Stable   Conditions'},   \newblock   {\em  Boundary-Layer   Meteorol.}
\textbf{90}, 495--520.

\bibitem[\protect\citeauthoryear{Hunt  et  al.}{1996}]{Huntetal} Hunt,
J. C. R., Shutts, G.  J., and Derbyshire, S.: 1996, \newblock {`Stably
Stratified Flows  in Meteorology'},  \newblock {\em Dyn.   Atmos.  and
Oceans} \textbf{23}, 63--79.

\bibitem[\protect\citeauthoryear{Kaimal  and  Finnigan}{1994}]{Kaimal}
Kaimal, J.  C.  and Finnigan,  J. J.: 1994, \newblock {\em Atmospheric
Boundary  Layer  Flows:   Their  Structure  and  Measurement},  Oxford
University Press, Oxford, UK, 289 pp.



\bibitem[\protect\citeauthoryear{Kosovi\'{c}                        and
Curry}{2000}]{Kosovic}  Kosovi\'{c},   B.   and  Curry   J.~A.:  2000,
\newblock {`A  Large Eddy Simulation  Study of a  Quasi-Steady, Stably
Stratified    Atmospheric    Boundary    Layer'},    \newblock    {\em
J. Atmos. Sci.} \textbf{57}, 1052--1068.


\bibitem[\protect\citeauthoryear{Lilly}{1967}]{Lilly}   Lilly,  D.~K.:
1967,  \newblock  {`The Representation  of  Small-Scale Turbulence  in
Numerical  Simulation  Experiments'},  in  \newblock {\em  Proc.   IBM
Scientific Computing Symposium on Environmental Sciences}, 195-210.

\bibitem[\protect\citeauthoryear{Lilly}{1992}]{Lilly2}  Lilly,  D.~K.:
1992, \newblock {`A Proposed Modification of the Germano Subgrid-scale
Closure Method'}, \newblock {\em Phys. Fluids A} \textbf{4}, 633--635.

\bibitem[\protect\citeauthoryear{Mahli}{1995}]{Mahli}   Mahli,  Y.~S.:
1995,  \newblock {`The  Significance of  the Dual  Solutions  for Heat
Fluxes  Measured  by  the  Temperature Fluctuation  Method  in  Stable
Conditions'},  \newblock {\em  Boundary-Layer  Meteorol.} \textbf{74},
389--396.

\bibitem[\protect\citeauthoryear{Mahrt}{1989}]{Mahrt89}   Mahrt,   L.:
1989, \newblock {`Intermittency of Atmospheric Turbulence'}, \newblock
{\em J. Atmos. Sci.} \textbf{46}, 79--95.

\bibitem[\protect\citeauthoryear{Mahrt}{1998a}]{Mahrt98}   Mahrt,  L.:
1998a,   \newblock  {`Stratified   Atmospheric  Boundary   Layers  and
Breakdown of  Models'}, \newblock  {\em Theoret. Comput.   Fluid Dyn.}
\textbf{11}, 263--279.

\bibitem[\protect\citeauthoryear{Mahrt}{1998b}]{Mahrt98b}  Mahrt,  L.:
1998b,  \newblock {`Flux  Sampling Errors  for Aircraft  and Towers'},
\newblock {\em J. Atmos. Oceanic Technol.} \textbf{15}, 416--429.



\bibitem[\protect\citeauthoryear{Mahrt   and  Vickers}{2002}]{Mahrt02}
Mahrt,  L. and  Vickers,  D.: 2002,  \newblock {`Contrasting  Vertical
Structures   of   Nocturnal    Boundary   Layers'},   \newblock   {\em
Boundary-Layer Meteorol.} \textbf{105}, 351--363.

\bibitem[\protect\citeauthoryear{Mason}{1989}]{Mason89}   Mason,   P.:
1989, \newblock {`Large-eddy  Simulation of the Convective Atmospheric
Boundary Layer'}, \newblock {\em J. Atm. Sci.} \textbf{46}, 1492-1516.

\bibitem[\protect\citeauthoryear{Mason}{1994}]{Mason} Mason, P.: 1994,
\newblock   {`Large-eddy  Simulation:   A  Critical   Review   of  the
Technique'}, \newblock {\em Q.  J. Roy.  Meteorol. Soc.} \textbf{120},
1--26.

\bibitem[\protect\citeauthoryear{Mason                              and
Derbyshire}{1990}]{MasonDerb}  Mason,  P.~J.   and Derbyshire,  S.~H.:
1990,  \newblock  {`Large-Eddy  Simulation  of  the  Stably-Stratified
Atmospheric Boundary Layer'}, \newblock {\em Boundary-Layer Meteorol.}
\textbf{53}, 117--162.

\bibitem[\protect\citeauthoryear{McNider    et    al.}{1995}]{McNider}
McNider, R.  T.,  England, D. E., Friedman, M. J.,  and Shi, X.: 1995,
\newblock {`Predictability of the Stable Atmospheric Boundary Layer'},
\newblock {\em J. Atmos. Sci.} \textbf{52}, 1602--1614.



\bibitem[\protect\citeauthoryear{Monin    and    Yaglom}{1971}]{Monin}
Monin, A. S. and Yaglom, A. M.: 1971, \newblock {\em Statistical Fluid
Mechanics: Mechanics of Turbulence}, Vol. 1, MIT Press, Cambridge, MA,
769 pp.

\bibitem[\protect\citeauthoryear{Nieuwstadt}{1984a}]{Nieuwstadt84a}
Nieuwstadt, F. T. M.: 1984a, \newblock {`Some Aspects of the Turbulent
Stable  Boundary  Layer'},  \newblock {\em  Boundary-Layer  Meteorol.}
\textbf{30}, 31--55.

\bibitem[\protect\citeauthoryear{Nieuwstadt}{1984b}]{Nieuwstadt84b}
Nieuwstadt, F.   T. M.: 1984b, \newblock {`The  Turbulent Structure of
the Stable, Nocturnal Boundary Layer'}, \newblock {\em J. Atmos. Sci.}
\textbf{41}, 2202--2216.

\bibitem[\protect\citeauthoryear{Nieuwstadt}{1985}]{Nieuwstadt85}
Nieuwstadt, F.  T.  M.: 1985,  \newblock {`A Model for the Stationary,
Stable  Boundary Layer'},  in J.   C. R.   Hunt (ed.),  \newblock {\em
Turbulence  and Diffusion  in Stable  Environments},  Clarendon Press,
Oxford, UK, 149--179.

\bibitem[\protect\citeauthoryear{Ohya}{2001}]{Ohya01}  Ohya, Y.: 2001,
\newblock  {`Wind-Tunnel Study of  Atmospheric Stable  Boundary Layers
over  a  Rough  Surface'},  \newblock {\em  Boundary-Layer  Meteorol.}
\textbf{98}, 57--82.

\bibitem[\protect\citeauthoryear{Ohya et al.}{1997}]{Ohya97} Ohya, Y.,
Neff,  D.   E., and  Meroney,  R.   N.:  1997, \newblock  {`Turbulence
Structure in  a Stratified  Boundary Layer under  Stable Conditions'},
\newblock {\em Boundary-Layer Meteorol.} \textbf{83}, 139--161.

\bibitem[\protect\citeauthoryear{Orszag     and    Pao}{1974}]{Orszag}
Orszag, S.~A., and Pao, Y.-H.: 1974, \newblock {`Numerical Computation
of   Turbulent   Shear  Flows'},   \newblock   {\em  Adv.    Geophys.}
\textbf{18A}, 224--236.

\bibitem[\protect\citeauthoryear{Pahlow et al.}{2001}]{Pahlow} Pahlow,
M.,  Parlange, M.  B.,  and Port\'{e}-Agel,  F.: 2001,  \newblock {`On
Monin-Obukhov Similarity  in the Stable  Atmospheric Boundary Layer'},
\newblock {\em Boundary-Layer Meteorol.} \textbf{99}, 225--248.

\bibitem[\protect\citeauthoryear{Piomelli   and  Liu}{1995}]{Piomelli}
Piomelli, U., and Liu,  J.: 1995, \newblock {`Large-eddy Simulation of
Rotating Channel  Flows using  a Localized Dynamic  Model'}, \newblock
{\em Phys. Fluids} \textbf{7}, 839--848.

\bibitem[\protect\citeauthoryear{Port\'{e}-Agel  et al.}{2000}]{Porte}
Port\'{e}-Agel, F., Meneveau, C., and Parlange, M. B.: 2000, \newblock
{`A   Scale-Dependent  Dynamic   Model   for  Large-Eddy   Simulation:
Application to a Neutral  Atmospheric Boundary Layer'}, \newblock {\em
J. Fluid Mech.} \textbf{415}, 261--284.

\bibitem[\protect\citeauthoryear{Port\'{e}-Agel}{2004}]{Porte04}
Port\'{e}-Agel, F.: 2004,  \newblock {`A Scale-Dependent Dynamic Model
for  Scalar Transport  in  LES of  the  Atmospheric Boundary  Layer'},
\newblock {\em Boundary-Layer Meteorol.} \textbf{112}, 81-105.

\bibitem[\protect\citeauthoryear{Poulos  et al.}{2002}]{CASES} Poulos,
G.   S., and  Coauthors: 2002,  \newblock {`CASES-99:  A Comprehensive
Investigation of the Stable Nocturnal Boundary Layer'}, \newblock {\em
Bull. Amer. Meteorol. Soc.} \textbf{83}, 555--581.

\bibitem[\protect\citeauthoryear{Revelle}{1993}]{Revelle}  ReVelle, D.
O.: 1993, \newblock {`Chaos and ``Bursting'' in the Planetary Boundary
Layer'}, \newblock {\em J. App.  Meteorol.}  \textbf{32}, 1169--1180.


\bibitem[\protect\citeauthoryear{Saiki  et  al.}{2000}]{Saiki}  Saiki,
E. M., Moeng, C.-H., and Sullivan, P. P.: 2000, \newblock {`Large-Eddy
Simulation  of  the  Stably  Stratified  Planetary  Boundary  Layer'},
\newblock {\em Boundary-Layer Meteorol.} \textbf{95}, 1--30.

\bibitem[\protect\citeauthoryear{Smedman}{1988}]{Smedman} Smedman, A.:
1988, \newblock  {`Observations of a  Multi-Level Turbulence Structure
in  a  Very  Stable   Atmospheric  Boundary  Layer'},  \newblock  {\em
Boundary-Layer Meteorol.} \textbf{44}, 231--253.


\bibitem[\protect\citeauthoryear{Sorbjan}{1989}]{Sorbjan} Sorbjan, Z.:
1989,  \newblock  {\em   Structure  of  Atmospheric  Boundary  Layer},
Prentice-Hall, Englewood Cliffs, NJ, 317 pp.

\bibitem[\protect\citeauthoryear{Stoll                              and
Port\'{e}-Agel}{2004}]{Stoll}  Stoll, R., and  Port\'{e}-Agel, F.:
2004, \newblock  {`Effect of Roughness on  Surface Boundary Conditions
for Large-eddy Simulation'},  \newblock {\em Boundary-Layer Meteorol.}
In Press.

\bibitem[\protect\citeauthoryear{Stull}{1988}]{Stull}  Stull,  R.  B.:
1988, \newblock  {\em An Introduction to  Boundary Layer Meteorology},
Kluwer Academic Publishers, Dordrecht, The Netherlands, 670 pp.

\bibitem[\protect\citeauthoryear{van de Wiel}{2002}]{vandeWiel} van de
Wiel, B.:  2002, \newblock {`Intermittent  Turbulence and Oscillations
in the Stable  Boundary Layer over Land'}, \newblock  {\em PhD Thesis,
Wageningen University, Netherlands}, 129 pp.

\bibitem[\protect\citeauthoryear{Vickers   and   Mahrt}{1997}]{Vick97}
Vickers, D. and Mahrt, L.:  1997, \newblock {`Quality Control and Flux
Sampling  Problems  for  Tower  and Aircraft  Data'},  \newblock  {\em
J. Atmos. Oceanic Technol.} \textbf{14}, 512--526.

\bibitem[\protect\citeauthoryear{Vickers   and   Mahrt}{2003}]{Vick03}
Vickers, D.  and  Mahrt, L.: 2003, \newblock {`The  Cospectral Gap and
Turbulent  Flux  Calculations'}, \newblock  {\em  J.  Atmos.   Oceanic
Technol.} \textbf{20}, 660--672.

\bibitem[\protect\citeauthoryear{Viterbo    et    al.}{1999}]{Viterbo}
Viterbo,  P., Beljaars, A.,  Mahfouf, J.-F.,  and Teixeira,  J.: 1999,
\newblock  {`The  Representation of  Soil  Moisture  Freezing and  its
Impact    on   the   Stable    Boundary   Layer'},    \newblock   {\em
Quart. J. Roy. Meteorol. Soc.} \textbf{125}, 2401--2426.


\bibitem[\protect\citeauthoryear{Wyngaard}{1973}]{Wyngaard}   Wyngaard,
J.  C.:  1973, \newblock  {`On Surface Layer  Turbulence'}, in  D.  A.
Haugen (ed.),  \newblock {\em Workshop  on Micrometeorology}, American
Meteorological Society, Boston, 109-149.

\bibitem[\protect\citeauthoryear{Zang  et al.}{1993}]{Zang}  Zang, Y.,
Street, R.~L.,  and Koseff, J.~R.:  1993, \newblock {`A  Dynamic Mixed
Subgrid-scale  Model and  its Application  to  Turbulent Recirculating
Flows'}, \newblock {\em Phys. Fluids. A} \textbf{5}, 3186--3196.

\end{thebibliography}
\end{document}